\documentclass[aps,prb,twocolumn,superscriptaddress]{revtex4-1}
\usepackage{physics}
\usepackage{graphicx}
\usepackage{amsmath}
\usepackage{amssymb}
\usepackage{bbm}
\usepackage{bm}
\usepackage{xcolor}
\usepackage[utf8]{inputenc}
\usepackage{ulem}

\newcommand{\up}{\uparrow}
\newcommand{\dn}{\downarrow}

\newcommand{\refeq}[1]{Eq.~(\ref{#1})}

\newcommand{\reffig}[1]{Fig.~\ref{#1}}
\newcommand{\refapp}[1]{Appendix~\ref{#1}}
\newcommand{\refsec}[1]{Sec.~\ref{#1}}
\newcommand{\eV}{\,\mathrm{eV}}
\newcommand{\mum}{\,\mathrm{\mu m}}
\newcommand{\nm}{\,\mathrm{nm}}
\newcommand{\K}{\,\mathrm{K}}
\newcommand{\fan}{F}
\newcommand{\lsco}{La$_{2-x}$Sr$_{x}$CuO$_4$}
\newcommand{\lbco}{La$_{2-x}$Ba$_{x}$CuO$_4$}

\newcommand{\ybco}{YBa$_2$Cu$_3$O$_{7-x}\,$}

\begin{document}
\title{Josephson lattice model for phase fluctuations of local pairs\\in copper-oxide superconductors}
\author{Malte~Harland}
\affiliation{Institute of Theoretical Physics, University of Hamburg, Jungiusstra{\ss}e 9, 20355 Hamburg, Germany}
\author{Sergey~Brener}
\affiliation{Institute of Theoretical Physics, University of Hamburg, Jungiusstra{\ss}e 9, 20355 Hamburg, Germany}
\author{Alexander~I.~Lichtenstein}
\affiliation{Institute of Theoretical Physics, University of Hamburg, Jungiusstra{\ss}e 9, 20355 Hamburg, Germany}
\author{Mikhail~I.~Katsnelson}
\affiliation{Institute for Molecules and Materials, Radboud University, 6525AJ, Nijmegen, the Netherlands}
\date{\today}
\begin{abstract}
  We derive an expression for the effective Josephson coupling from the microscopic Hubbard model. It serves as a starting point for the description of phase fluctuations of local Cooper pairs in $d_{x^2-y^2}$-wave superconductors in the framework of an effective $XY$ model of plaquettes, the Josephson lattice. The expression for the effective interaction is derived by means of the local-force theorem, and it depends on local symmetry-broken correlation functions that we obtain using the cluster dynamical mean-field theory. Moreover, we apply the continuum limit to the Josephson lattice to obtain an expression for the gradient term in the Ginzburg-Landau theory and compare predicted London penetration depths and Kosterlitz-Thouless transition temperatures with experimental data for \ybco.
\end{abstract}
\maketitle

\section{Introduction}
Since the discovery of High-$T_c$ superconductivity\cite{Bednorz1986} many types of competing orders have been considered\cite{Anderson1987,Dagotto1994,Lee2006,Scalapino2012,Demler2004,Berg2007,Fradkin2015,Keimer2015} which could have strong effects on the superconducting critical temperature. It is generally recognized that in the underdoped copper-oxide superconductors the Kosterlitz-Thouless (KT) physics \cite{Kosterlitz1973} is crucial due to strong phase fluctuations \cite{Uemura1989,Emery1995,Alvarez1996,Kwon2001,Sharapov2001,Benfatto2004}. Important progress in the non-perturbative\cite{Imada1998} treatment of the antiferromagnetism and $d$-wave superconductivity (dSC) in the Hubbard model is related to the cluster dynamical mean-field theory (CDMFT)\cite{Lichtenstein2000,Kotliar2001,Maier2005,Haule2007,Civelli2008,Ferrero2009,Sordi2012,Gull2013,Hebert2015,Fratino2016,Harland2016,Braganca2018,Wu2018}. It yields a local d-wave superconducting order parameter, but it neglects spatial correlations beyond the cluster. Recently, large scale DMRG calculations \cite{Jiang2018,Jiang2018a} confirmed the existence of long-range superconducting correlations in the Hubbard and $t-J$ models. The CDMFT prediction for the superconducting critical temperature $T_c$, however,  is too high, and long-range corrections are required for a realistic description.

In this work, we apply a truncated description, coarse graining, which is a very general and powerful tool that allows for a replacement of a microscopic by a macroscopic description with microscopically defined parameters. The prototype procedure in the theory of magnetism has opened the way to a quantitative theory of magnetism for real materials\cite{Liechtenstein1987,Katsnelson2000,Eriksson2017}. We map the CDMFT solution of the Hubbard model onto the Josephson lattice model assuming a separation of energy scales that correspond to the dSC phase (Goldstone) and amplitude (Higgs) fluctuations. We start from a numerically exact solution of the minimal CDMFT problem with the two-by-two plaquette in a superconducting bath as an effective impurity, and we obtain a local cluster dSC order parameter. Subsequently, we introduce long-range perturbations in the dSC-phase and derive the effective coupling of the Josephson lattice model that describes phase fluctuations. 

\section{Theory: From Hubbard to Josephson}
The one-band Hubbard model\cite{Hubbard1963}, which is widely accepted to capture the essential physics of cuprates\cite{Dagotto1994,Lee2006,Scalapino2012}, reads
\begin{equation}
  \label{eq:Hubbard}
  H = -\sum_{k\sigma } t(k) c_{k\sigma}^{\dag} c_{k\sigma } + U \sum_r n_{r\up } n_{r\dn },\\
\end{equation}
where $t(k)$ are the Fourier-transformed hopping parameters and $U$ is the interelectron Coulomb repulsion parameter on site $r$. $c_{r\sigma }^\dag$ and $c_{r\sigma }$, ($c_{k\sigma }^{\dag}$ and $c_{k\sigma }$) are electron creation and annihilation operators in site (momentum) representation, respectively, and $n_{r\sigma }=c_{r\sigma }^\dag c_{r\sigma }$. We use the nearest neighbor hopping of the square lattice $|t|$ as energy unit and for the next-nearest neighbor hopping $t'/t = -0.3$ for \ybco\cite{Pavarini2001}.

In principle, the description of the two-dimensional (2D) square lattice defined by the dispersion
\begin{equation}
  \label{eq:tk2d}
  t^{2D}(k)= 2t \left( \cos(k_x) + \cos(k_y) \right) + 4t^\prime \cos(k_x) \cos(k_y),
\end{equation}
is sufficient to obtain local pairs within the strong-coupling planes. However, in order to calculate an effective interlayer Josephson coupling and the out-of-plane London penetration depth, it is essential to have interlayer hopping. Our three-dimensional (3D) calculations, that include interlayer hopping, use an anisotropic infinite layer model\cite{Chakravarty1993,Andersen1995} with the dispersion
\begin{equation}
  \label{eq:tk3ds}
  t^{3D}(k) = t^{2D}(k) + 2 \frac{t_\perp}{4} \left( \cos(k_x) - \cos(k_y) \right)^2 \cos(k_z),
\end{equation}
which has interlayer hopping of $d_{x^2-y^2}$ symmetry and is generic for cuprates. For \refeq{eq:tk2d} and \refeq{eq:tk3ds}, $k_x$, $k_y$ and $k_z$ are in the Brillouin zone. Note that below we introduce a two-by-two cluster formulation that corresponds to the reduced Brillouin zone (\refapp{app:tightbinding}). This requires the choice of unit lengths $a_a, a_b, a_c$ = $2\times 3.82$\,\AA, $2\times 3.82$\,\AA, $3.89$\,\AA\, that is twice the copper distance within the copper planes of YBCO\cite{Wu1987,Cava1987}. Further, we choose the simplified effective hopping of $t_\perp/t = 0.15$ for YBCO and the effective tight-binding hopping $|t|=0.35\eV$\cite{Andersen1995,Fratino2016}. The screened Coulomb interaction is set to a standard value, $U=8|t|$, of the order of the bandwidth.

\begin{figure}
  \centering
  \includegraphics{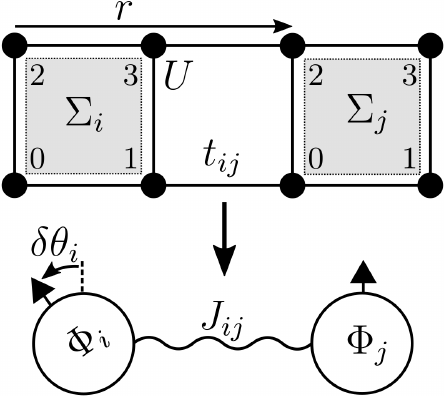}
  \caption{Illustration of the Hubbard-plaquette lattice ($t_{ij}$, $U$) with lattice vector $r$, self-energies $\Sigma_i$ and plaquette sites $0...3$. It is mapped to the Josephson lattice model with effective  coupling $J_{ij}$ of plaquettes due to phase fluctuations $\delta \theta_i$ of the $d$-wave superconducting order parameter $\Phi_i$.}
  \label{fig:illu}
\end{figure}

To address the specific problem of Josephson coupling in cuprates, we consider a local $U(1)$ rotation that changes the phase of the plaquette's dSC order parameter, similar to a rotation of an effective moment attributed to a two-by-two plaquette and keeps the amplitude of the local order parameter constant, see \reffig{fig:illu}. We investigate macroscopic phase coherence between the plaquettes, reminiscent of the description of magnetic ordering in terms of an effective Heisenberg Hamiltonian\cite{Liechtenstein1987,Katsnelson2000}. The model, that can address the issue of superconducting phase ordering, is the Josephson lattice model
\begin{equation}
H_{\mathrm{eff}}=-\sum_{ij}J_{ij}\cos \left( \theta _{i}-\theta _{j}\right),
\label{eq:hjosephson}
\end{equation}
i.e. an effective $XY$ model of plaquettes. $i,j$ are plaquette indices, and $\theta_i$ is the phase of the order parameter of plaquette $i$. The principal goal of our work is to obtain the Josephson coupling parameters $J_{ij}$ based on the Hubbard model solution of the well-established CDMFT\cite{Kotliar2001,Haule2007,Civelli2008,Hebert2015,Fratino2016,Harland2016,Braganca2018}. We consider the elementary plaquette in the copper layer as a supersite and introduce a superspinor $C_{i}^{\dag}=\left(c_{i\alpha }^{\dag}\right)$, where $i$ is the index of the plaquette and $\alpha = 0...3$ labels the sites within the plaquette, see \reffig{fig:illu}.  In order to describe the superconducting state, we use the Nambu-Gor'kov spinor representation of the Green function, which is a 2$\times $2 matrix. Thus, the full lattice Green function $G_{ij}$ is an $8\times 8$ matrix.

The explicit microscopic expressions of $J_{ij}$ is derived by calculating the microscopic variation of the thermodynamic potential $\Omega$ of the system under small variations of the dSC phases, and comparing the result with \refeq{eq:hjosephson}. $\Omega$ depends on the lattice Green function that we can express via the Dyson equation
\begin{align}
  \label{eq:senambu}
  \begin{pmatrix}
    G^{p\up} & F \\
    F  & G^{h \dn}
  \end{pmatrix}^{-1}_{ij}
                =
                \begin{pmatrix}
                  G^{p\up}_0  & 0 \\
                  0 & G^{h\dn}_0
                \end{pmatrix}^{-1}_{ij}
                    -
                    \delta_{ij}
                    \begin{pmatrix}
                      \Sigma^{p\up}  & S  \\
                      S  & \Sigma^{h \dn}
                    \end{pmatrix}_{i},
\end{align}
Where the last term is the local self-energy of the CDMFT (\refapp{app:cdmft}). The superscripts $p$ and $h$ denote particle and hole components of the Nambu-Gor'kov representation, respectively. The anomalous parts of the self-energy $S$ and Green function $F$ are matrices in plaquette sites $\alpha$ and describe local dSC pairing via the order parameter $\Phi_{\mathrm{dSC}}^{\mathrm{CDMFT}} = 2T\Tr_{\omega}F_{01}$ with $F_{01} = -F_{02}$, according to $d$-wave symmetry\cite{Lichtenstein2000}. $G_0$ denotes the non-interacting lattice Green function. Furthermore, we consider finite temperatures $T$, and, therefore, the correlation functions depend on fermionic Matsubara frequencies. The last term of \refeq{eq:senambu}, the local self-energy $\Sigma_i$, is obtained exactly by the numerical\cite{Parcollet2015,Seth2016,Gull2011} solution of the CDMFT.

In order to find the variation of the free energy
\begin{gather}
  \label{eq:lft}
  \Omega = \Omega _{sp}-\Omega _{dc},\nonumber\\
  \Omega _{sp} = -\Tr\ln \left( -G^{-1}\right),\\
  \Omega _{dc} = \Tr\Sigma G-\Phi,\nonumber
\end{gather}
with the Luttinger-Ward functional\cite{Luttinger1960} $\Phi$, we use the local-force theorem\cite{Katsnelson2000,Stepanov2019}
\begin{equation}
  \delta \Omega \simeq \sum_{ij} \Tr \left( \delta_{ij} G_{ii}\delta^\ast\Sigma_i 
  + \frac{1}{2} G_{ij} \delta^\ast\Sigma_j G_{ji} \delta^\ast\Sigma_i\right)\label{eq:omegavariation},
\end{equation}
where $\delta^\ast$ denotes the local variation of the self-energy $\Sigma$ without taking into account its variation due to the CDMFT self-consistency, and $G$ is the CDMFT Green function without variation. We omit matrix indices of intra-plaquette and Nambu space for simplicity. \refeq{eq:omegavariation} is rigorous in the first order of the phase variations $\delta\theta _{i}$\cite{Katsnelson2000}. However, we will use it also for the second order terms since the first order variation around the colinear state, $\theta_i =$ const., vanishes analytically (\refapp{app:gi}). It corresponds to neglecting vertex corrections\cite{Luttinger1960} that is reasonable to assume for the locally ordered phase with a well-pronounced, local order parameter\cite{Stepanov2018}. Thus, near the transition, it can be used as an estimate only.

We design the variation as an infinitesimal change of the local phase $\delta \theta_i$ in a homogeneous environment. Therefore, it reads
\begin{align}
  \label{eq:variationsigma}
  \begin{split}
    \delta^\ast\Sigma_i &= e^{i\delta\theta_i \sigma_z/2} \Sigma_i e^{-i\delta\theta_i \sigma_z/2} -\Sigma_i\\
    &=\begin{pmatrix} \Sigma^{p\up}_i && e^{i\delta\theta_i}S_i \\ e^{-i\delta\theta_i}S_i && \Sigma^{h\dn}_i \end{pmatrix} - \Sigma_i\\
    &\simeq \begin{pmatrix} 0 && \left(i\delta\theta_i - \frac{\left(\delta\theta_i\right)^2}{2}\right)S_i \\ \left(-i\delta\theta_i - \frac{\left(\delta\theta_i\right)^2}{2}\right)S_i && 0\end{pmatrix},
  \end{split}
\end{align}
in that the third Pauli matrix $\sigma_z$ acts in the Nambu-space. This variation affects only the phases of the anomalous part of the local self-energy. We substitute \refeq{eq:variationsigma} into \refeq{eq:omegavariation} and the two terms of the sum become
\begin{widetext}
  \begin{gather}
  \label{eq:je5}
  G_{ii} \delta^\ast\Sigma_{i} = 
  \begin{pmatrix}
    F_{ii} S_i \left(-i\delta\theta_i-\frac{\left(\delta\theta_i\right)^2}{2}\right) &&  G^{p\up}_{ii} S_i \left(i\delta\theta_i-\frac{\left(\delta\theta_i\right)^2}{2}\right) \\
    G^{h\dn}_{ii} S_i \left(-i\delta\theta_i-\frac{\left(\delta\theta_i\right)^2}{2}\right)  && F_{ii} S_i \left(i\delta\theta_i-\frac{\left(\delta\theta_i\right)^2}{2}\right)
  \end{pmatrix}
\end{gather}
\begin{gather}
  \label{eq:je6}
  G_{ij}\delta^\ast\Sigma_{j} G_{ji}\delta^\ast\Sigma_{i} = 
  \begin{pmatrix}
    -F_{ij}S_j F_{ji}S_i + G^{p\up}_{ij}S_j  G^{h\dn}_{ji}S_i  && \cdots\\
    \cdots && -F_{ij}S_j F_{ji}S_i + G^{h\dn}_{ij}S_j G^{p\up}_{ji}S_i
  \end{pmatrix}
\times  \delta\theta_i  \delta\theta_j
\end{gather}
We keep terms up to second order in $\delta\theta$, and since we are interested in the trace, we omit off-diagonals in \refeq{eq:je6}. \refeq{eq:je5} shows clearly that the trace makes the first order vanish. Using $\delta\theta_{ij} \equiv \left(\delta\theta_i - \delta\theta_j\right)$ and $2 \delta\theta_i \delta\theta_j = - \delta\theta^2_{ij} + \delta\theta^2_i + \delta\theta^2_j$, we can separate local and non-local phase variations,
\begin{equation}
  \label{eq:je8}
  \delta\Omega = \sum_{ij} \Tr_{\omega \alpha} \left(G^{p\up}_{ij}S_jG^{h\dn}_{ji}S_i - \delta_{ij}  F_{ii}S_i- F_{ij}S_jF_{ji}S_i\right)\delta\theta^2_i
  +\frac{1}{2}\sum_{ij} \Tr_{\omega \alpha} \left(F_{ij}S_jF_{ji}S_i  - G^{p\up}_{ij}S_jG^{h\dn}_{ji}S_i\right)\delta\theta^2_{ij}.
\end{equation}
\end{widetext}
The trace goes over Matsubara frequencies and over the sites within the plaquette ($\alpha$). Furthermore, the matrices form matrix-products in the $\alpha$-space whereas they are diagonal in Matsubara frequencies. In order to obtain \refeq{eq:je8} we have also used the lattice symmetry $G_{ij} =  G_{ji}$.

The term $\propto \delta\theta^2_i$ vanishes which reflects the gauge invariance of the theory (\refsec{app:gi}). The remaining term is that of only non-local phase fluctuations $\propto \delta\theta^2_{ij}$
\begin{equation}
  \label{eq:lft4}
  \delta \Omega \equiv \frac{1}{2} \sum_{ij} J_{ij} \delta \theta^2_{ij}
\end{equation}
which by comparison with \refeq{eq:hjosephson} defines $J_{ij}$. Thereby, we obtain the following expression of the Josephson lattice parameters
\begin{align}
  \label{eq:j}
  J_{ij}=T \Tr_{\omega \alpha} \left(- G_{ij}^{p\up}  S_{j} G_{ji}^{h\dn} S_{i} + F_{ij} S_{j} F_{ji}  S_{i} \right),
\end{align}
which is essentially the main result of the present work.

\section{Short-range Josephson lattice parameters\label{sec:j}}
\begin{figure}
  \centering
  \includegraphics{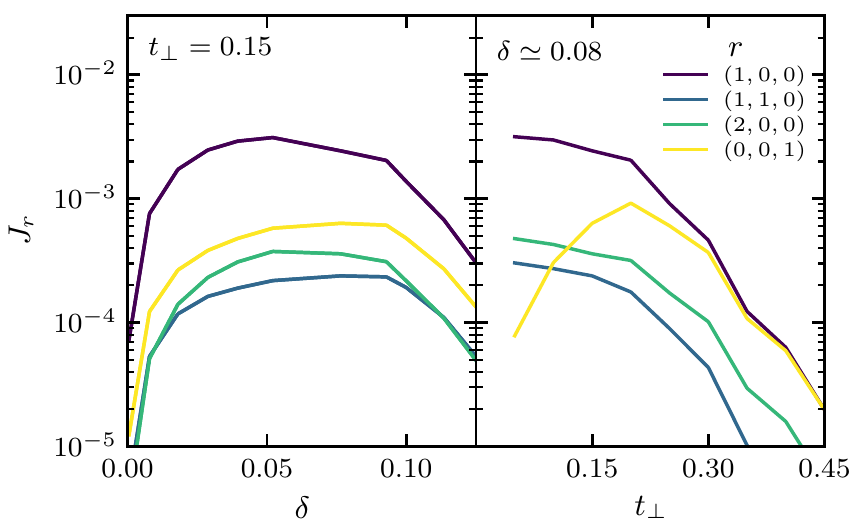}
  \caption{Josephson coupling $J_r$ as a function of doping $\delta$ (left) and interlayer hopping $t_\perp$ (right) for different plaquette translations $r$ at $T=1/52$}
  \label{fig:j}
\end{figure}
Effective Josephson couplings have been applied to investigate experiments in that interplane Josephson coupling has an essential role\cite{Okamoto2016,Okamoto2017}. We present a selection of the Josephson couplings $J_r$ for plaquette-translations $r$ in \reffig{fig:j}. $J_r$ reduces sharply with increasing plaquette-translation length $|r|$, and thus the short-range components of $J_r$ alone can give a complete description. The strongest coupling is $J_{100}$, followed by the interlayer coupling $J_{001}$. They have their maxima around $\delta = 0.05$ and $\delta = 0.1$, respectively. All couplings diminish at large dopings, $\delta > 0.1$. We observe in \refsec{sec:i} that this stems from the diminishing of the local orderparamter (amplitude) of the dSC.

In the range up to $t_\perp = 0.45$, $t_\perp$ has a diminishing effect on all in-plane $J_r$, shown in \reffig{fig:j} (right). In contrast, the interlayer coupling has to increase at small $t_\perp$ since there has to be $J_{001}=0$ in a system of disconnected layers ($t_\perp = 0$). $J_{001}$ becomes the second largest coupling at $t_\perp = 0.15$, and at $t_\perp = 0.2$ it reaches a maximum. For larger $t_\perp$ all couplings decrease, similar to the behavior at large dopings.

\begin{figure}
  \centering
  \includegraphics{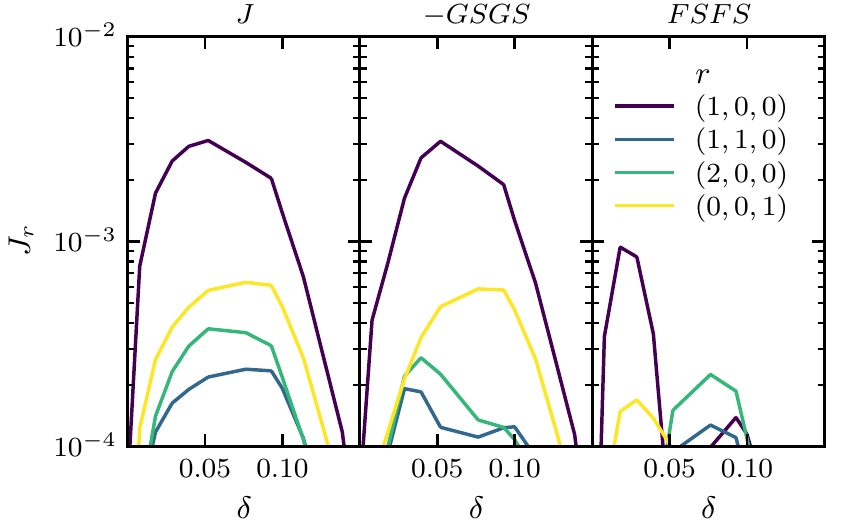}
  \caption{Josephson coupling $J_r$ (left) and its constituents, $GSGS$ (center) and $FSFS$ (right), as functions of doping $\delta$ and for different plaquette translations $r$ at $T=1/52\sim 0.02$, $t_\perp = 0.15$.}
  \label{fig:jgsgsfsfs}
\end{figure}
The first term of \refeq{eq:j} ($GSGS$) is negative, and the second ($FSFS$) is positive. $GSGS$ is a mixed term with normal ($G$) and anomalous ($S$) contributions. It makes the main contribution to $J$, see \reffig{fig:jgsgsfsfs}. $J$ can be finite only if there is a superconducting gap and therefore a finite anomalous self-energy $S$ as both terms depend on it. Regarding the largest contributions to the nearest neighbour Josephson coupling $J_{(1,0,0)}$, $GSGS$ is about 3 times as large as $FSFS$. However, at small dopings both terms contribute with similar magnitude, but their doping dependence can be very different. At $\delta \sim 0.05$ the first term drops sharply and $J_{(1,0,0)}$ is defined by $GSGS$. The second and third in-plane nearest neighbors have contributions from both terms and they can be of similar magnitude. However, the doping dependence have different local features, e.g. a local minimum of the second term appears in $J_{(1,1,0)}$, at a point where the first term has a maximum.

\section{Superconducting stiffness\label{sec:i}}
\begin{figure}
  \centering
  \includegraphics{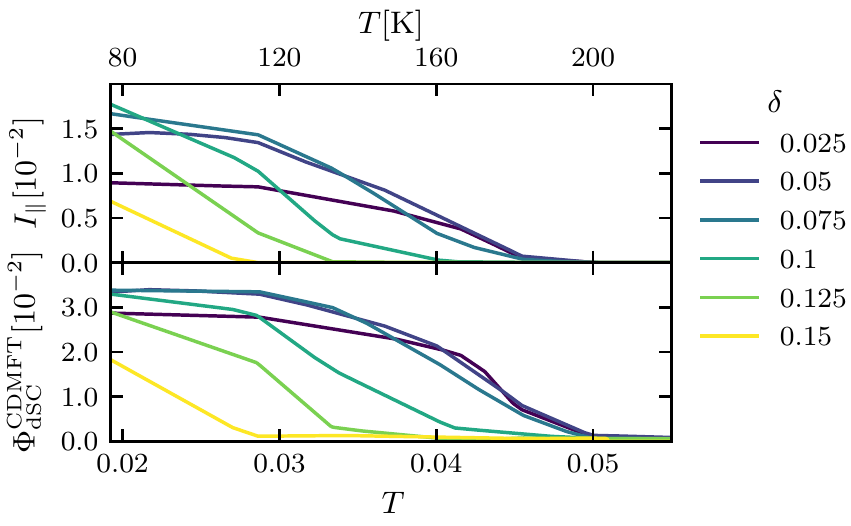}
  \caption{Superconducting stiffness $I_{\parallel}$ (top) and order parameter for local Cooper-pair formation $\Phi^{\mathrm{CDMFT}}_{\mathrm{dSC}}$ (bottom) as functions of the temperature $T$ for various dopings $\delta$ ($t_\perp = 0$).}
  \label{fig:stifftpd}
\end{figure}
In order to study macroscopic observables of the Josephson lattice model, we take the continuum, long-wavelength limit of \refeq{eq:hjosephson}. In this limit, the interaction becomes the superconducting stiffness (\refapp{app:tl})
\begin{gather}
  \label{eq:stiffaak}
    I_{ab} = \frac{T}{\left(2\pi\right)^d}\int\! d^dk \Tr_{\omega \alpha}\\
    \times \left( -\frac{\partial G^{p\up}(k)}{\partial k_a} S \frac{\partial G^{h\dn}(k)}{\partial k_b} S
     +\frac{\partial F(k)}{\partial k_a} S \frac{\partial F(k)}{\partial k_b} S\right)\nonumber
\end{gather}
with the effective Hamiltonian
\begin{equation}
  \label{eq:hjosephsoncontinuum}
  H_{\mathrm{eff}}=\frac{1}{2}\sum_{ab} I_{ab}\int d^dr\,\frac{\partial \theta }{\partial r_{a}}\frac{\partial \theta }{\partial r_{b}}.
\end{equation}
For our model $I_{ab}$ consists of an in-plane $I_\parallel$ and a perpendicular $I_\perp$ component. $I_\perp$ is non-zero only in the (3D) case of interlayer hoppings $t_{\perp} > 0$. \refeq{eq:hjosephsoncontinuum} can be viewed as the limit of the general Ginzburg-Landau equation for the case of a constant absolute value of the superconducting order parameter and negligible electromagnetic fields. The latter condition is controlled by slow spatial variations of the phase of the order parameter.

We start the discussion of the dSC stiffness for the 2D case of $t_{\perp} = 0$. The temperature dependence of the dSC stiffness can be divided into two, qualitatively different, regions depending on the hole-dopings of the copper planes $\delta$, see \reffig{fig:stifftpd} (top). In the underdoped regime ($0 \lesssim \delta \lesssim 0.075$) the temperature at that $I_\parallel$ becomes non-zero is constant. Furthermore, $I_\parallel$ shows saturation with decreasing $T$ only in the underdoped regime. In contrast, in the optimal- to over-doped regime ($0.1 \lesssim \delta \lesssim 0.15$), the temperature at that $I_\parallel$ becomes non-zero, as well as the low-temperature ($T\sim 0.02$) value of $I_\parallel$, decrease with larger doping. The low-temperature doping dependence of $I_\parallel$ qualitatively agrees with experimental studies on YBCO\cite{Uemura1988,Boyce2000} (and \lsco\cite{Panagopoulos1999}) and also with a study of the intensity of a current-current correlation function's Drude-like peak\cite{Haule2007}. Note, that the latter method can give just a number for the superfluid density whereas our approach allows to restore the whole Hamiltonian with the non-local effective Josephson parameters.

Regarding the accuracy of the local-force theorem, it is important to check whether the saturation of the local order parameter $\Phi_{\mathrm{dSC}}^{\mathrm{CDMFT}}$ with respect to decreasing temperature is reached. If this is the case, the the phase fluctuations are effectively decoupled from the Higgs mode and can be considered independently. Otherwise, amplitude fluctuations of the dSC can become stronger and vertex corrections, that we neglect, become significant\cite{Stepanov2018}. Our calculations show a saturation of $\Phi_{\mathrm{dSC}}^{\mathrm{CDMFT}}$ at $T\sim 0.02$ for dopings $\delta \lesssim 0.1$. Arbitrary low temperatures can not be reached because of the CTQMC-fermionic sign problem\cite{Gull2011}.

\begin{figure}
  \centering
  \includegraphics{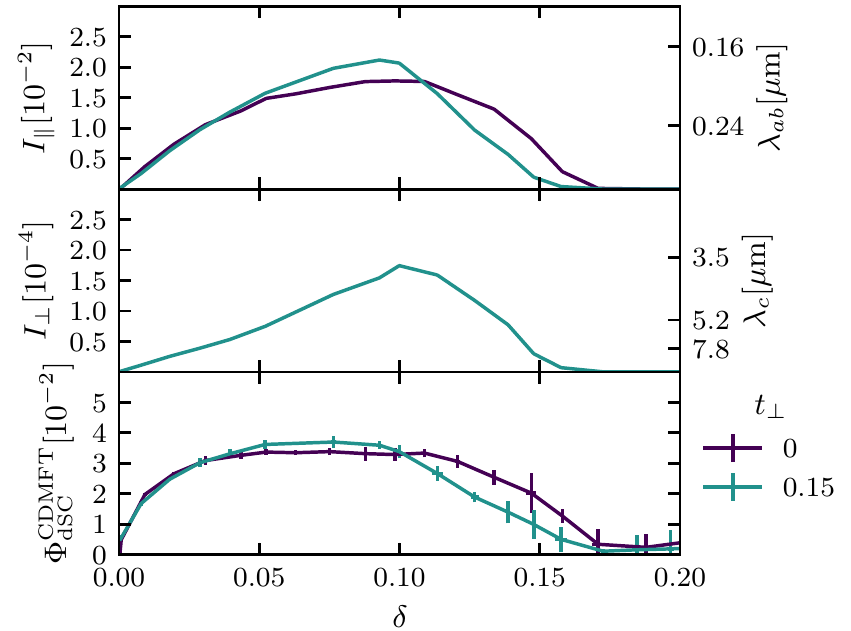}
  \caption{In-plane superconducting stiffness $I_{\parallel}$ (top, left), in-plane penetration depth $\lambda_{ab}$ (top, right), perpendicular superconducting stiffness $I_{\perp}$ (center, left), perpendicular penetration depth (center, right) and CDMFT dSC order parameter $\Phi^{\mathrm{CDMFT}}_{\mathrm{dSC}}$ (bottom) as functions of doping $\delta$ at $T = 1/52$ for different interlayer hoppings $t_\perp$.}
  \label{fig:stiff_sco_eps_dop}
\end{figure}
In \reffig{fig:stiff_sco_eps_dop} we compare the in-plane/perpendicular dSC stiffness and penetration depth as well as the order parameter of local Cooper pair formation for different $t_\perp$ (3D). $t_\perp$ has a minor impact on $I_\parallel$ which is probably related to our special choice of in-plane plaquette and to the mean-field character of the CDMFT. The perpendicular hopping $t_\perp = 0.15$ enhances $I_\parallel$ at optimal doping ($\delta \sim 0.1$) and reduces $I_\parallel$ at overdoping. At small dopings ($\delta < 0.05$), $I_\parallel$ is almost independent of $t_\perp$. Furthermore, for $t_\perp = 0.15$, $I_\parallel$ is two orders of magnitude larger than $I_\perp$ (\reffig{fig:stiff_sco_eps_dop}, center) reflecting the fact that, according to the Josephson lattice model, the superfluid is more concentrated within the strongly coupled copper planes. A comparison of $I_{\parallel/ \perp}$ with $\Phi_{\mathrm{dSC}}^{\mathrm{CDMFT}}$ (\reffig{fig:stiff_sco_eps_dop}, bottom) shows that $I_{\parallel/ \perp}$ has a more pronounced dome shape whereas $\Phi_{\mathrm{dSC}}^{\mathrm{CDMFT}}$ has a plateau, up to almost half-filling. Thus, relative to $\Phi_{\mathrm{dSC}}^{\mathrm{CDMFT}}$ the profile of $I_{\parallel/ \perp}$ is suppressed in the underdoped regime.

$I$ is closely related to the London penetration depth\cite{Emery1995,Sachdev2011} (\refapp{app:lpd}), i.e.
\begin{equation}
  \label{eq:lpd}
  \lambda^{-2} = \frac{16\pi e^2}{\hbar^2 c^2} I.
\end{equation}
$\lambda$ has been measured in several experiments on \ybco, also at different oxygen dopings $x$. The low-temperature values lie in the range of $\lambda_{ab} = 0.1 - 0.24\mum$ and $\lambda_c = 0.6-7.8\mum$\cite{Basov1995,Homes1995,Homes1999,Liu1999,Pereg-Barnea2004,Homes2004,Dordevic2013}. Finite temperature effects can add $\Delta \lambda_{ab} \sim 0.1\mum$ around $T\sim 80\K$\cite{Kamal1998}. In the underdoped region ($x=0.4$), the penetration depth is $\lambda_{ab} = 0.24\mum$ which is within the predicted range by our theory, around $\delta \sim 0.03$. Note, that the relation between the oxygen doping of YBCO $x$ and the hole doping of the copper-oxide planes $\delta$ is understood only qualitatively. Our largest value of $\lambda_{ab} \sim 0.16\mum$ is similar to the experimental result of $\lambda_{ab} = 0.15\mum$ for $x=0.05$ (optimal oxygen doping). Regarding the $c$-direction for the underdoped regime ($x=0.3 - 0.5$), experiments have found $\lambda_c = 5.2 - 7.8\mum$ which we have calculated around $\delta = 0.025 - 0.05$. In our calculations $\lambda$ is very sensitive to the details of the electronic interlayer properties (\refapp{app:ihop}) and the uncertainty in the interlayer hopping limits the accuracy of our predictions of $\lambda$.

\begin{figure}
  \centering
  \includegraphics{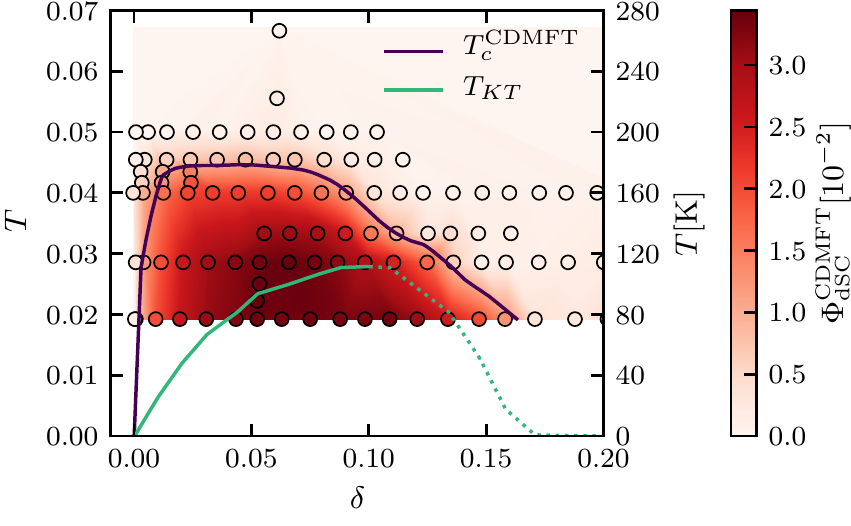}
  \caption{Phase diagram of the local dSC order parameter $\Phi^{\mathrm{CDMFT}}_{\mathrm{dSC}}$ with critical temperature $T^{\mathrm{CDMFT}}_c$ depending on the temperature $T$ and doping $\delta$ ($t_\perp = 0$).  Circles denote CDMFT calculations. The transition temperature of the Josephson lattice model $T_{KT}$ has been calculated from the superconducting stiffness at $T = 1/52$, at that $I_\parallel(T)$ is (not) saturated for the solid (dotted) part.}
  \label{fig:sco_dop_temp}
\end{figure}
In 2D, the $XY$ model of \refeq{eq:hjosephson} exhibits the KT transition that corresponds to the unbinding of vortex-antivortex pairs. The transition temperature reads\cite{Nelson1977}
\begin{equation}
  \label{eq:tkt}
  T_{KT} = \frac{\pi}{2} I_{\parallel}
\end{equation}
This proportionality of transition temperature and dSC stiffness can explain the Uemura relation\cite{Uemura1989} that has been measured in underdoped copper-oxides, via the muon spin relaxation rate. At $T<T_{KT}$ there is no real long-range order in the system but power-law decay of the correlation function of the superconducting order parameter. In this sense, interlayer tunneling is essentially important to allow for a dimensional crossover and long-range order\cite{Mermin1966,Benfatto2007}. In \reffig{fig:sco_dop_temp} we present the transition temperatures of the CDMFT $T_c^{\mathrm{CDMFT}}$, i.e. of local pair formation, and of the KT transition $T_{KT}$. We use $I$ of the lowest temperature available, $T\sim 0.02$, to calculate the KT transition temperature. At $\delta\lesssim 0.1$ the low-temperature saturation of $\Phi^{\mathrm{CDMFT}}_{\mathrm{dSC}}$ and $I_\parallel$ has been reached (\reffig{fig:stifftpd}), and thus, the application of our method is reliable. At $\delta\gtrsim 0.1$ amplitude fluctuations can change the transition temperature.

The suppression of the dSC by phase fluctuations is most pronounced at small dopings. This is where local Cooper-pairs, according to CDMFT, are well defined, up to half-filling. At half-filling the system is a Mott insulator\cite{Sordi2010,Sordi2012,Fratino2016} (\refapp{app:cdmft}), for which we have added a $T=0$ data point of prior CDMFT studies\cite{Kancharla2008}. The case of $T^{\mathrm{CMDFT}}_c > T_{KT}$ suggests a pseudogap interpretation of preformed meta-stable pairs\cite{Emery1995,Chen2005} in the underdoped copper-oxides. However, CDMFT supports other explanations as well\cite{Civelli2008,Sordi2012,Gull2013,Harland2016}. Note, that local antiferromagnetic fluctuations are included by CTQMC, but antiferromagnetic ordering and long-ranged spin waves are not. The latter can contribute to the suppression of superconductivity in cuprates, particularly  at $\delta \lesssim 0.05$\cite{Damascelli2003}. The maximum transition temperature of CDMFT is $T_c^{\mathrm{CDMFT},\,\mathrm{max}} \sim 180\K$, that is nearly twice as large than the experimental value\cite{Wu1987}. In contrast, including phase fluctuations gives a major correction, as $T_{KT}^{\mathrm{max}}\sim 120\K$. A comparison with the critical temperature of YBCO $T_c = 93K$\cite{Homes2004} and its Nernst region, that extends over a range up to $20\K$\cite{Zeh1990,Ri1994,Wang2006}, shows that the Josephson lattice model and phase disorder can be important for a quantitative description.

\section{Conclusion}
We have derived a mapping from the Hubbard to the Josephson lattice model, i.e. \refeq{eq:j}, and obtained effective couplings that will be interesting to study further in a more realistic bilayer model for e.g. \ybco or \lbco\cite{Kivelson2003,Berg2007,Wollny2009,Rajasekaran2018} in particular in the framework of the $XY$ model. At $T\sim t/50$ our theory is applicable to the underdoped regime as there the order parameter is well defined and the assumption of the separation of energy scales of amplitude and phase fluctuations is reasonable. Further, we have used analytical results of the $XY$ model to compare predictions, based on the obtained effective couplings, to experiments on \ybco. The London penetration depths have been confirmed to be reasonable estimates, and the KT transition lies closer to the experimental value than the critical temperature of the CDMFT which can indicate long-range phase disorder effects.

\begin{acknowledgments}
We thank G.~Homann, L.~Mathey and A.~Millis for discussions. MH, SB and AIL acknowledge support by the Cluster of Excellence 'Advanced Imaging of Matter' of the Deutsche Forschungsgemeinschaft (DFG) - EXC 2056 - project ID 390715994 and by the DFG SFB 925. MIK acknowledges a financial support from NWO via Spinoza Prize. The computations were performed with resources provided by the North-German Supercomputing Alliance (HLRN).
\end{acknowledgments}

\bibliography{josephson}

\begin{thebibliography}{75}%
\makeatletter
\providecommand \@ifxundefined [1]{%
 \@ifx{#1\undefined}
}%
\providecommand \@ifnum [1]{%
 \ifnum #1\expandafter \@firstoftwo
 \else \expandafter \@secondoftwo
 \fi
}%
\providecommand \@ifx [1]{%
 \ifx #1\expandafter \@firstoftwo
 \else \expandafter \@secondoftwo
 \fi
}%
\providecommand \natexlab [1]{#1}%
\providecommand \enquote  [1]{``#1''}%
\providecommand \bibnamefont  [1]{#1}%
\providecommand \bibfnamefont [1]{#1}%
\providecommand \citenamefont [1]{#1}%
\providecommand \href@noop [0]{\@secondoftwo}%
\providecommand \href [0]{\begingroup \@sanitize@url \@href}%
\providecommand \@href[1]{\@@startlink{#1}\@@href}%
\providecommand \@@href[1]{\endgroup#1\@@endlink}%
\providecommand \@sanitize@url [0]{\catcode `\\12\catcode `\$12\catcode
  `\&12\catcode `\#12\catcode `\^12\catcode `\_12\catcode `\%12\relax}%
\providecommand \@@startlink[1]{}%
\providecommand \@@endlink[0]{}%
\providecommand \url  [0]{\begingroup\@sanitize@url \@url }%
\providecommand \@url [1]{\endgroup\@href {#1}{\urlprefix }}%
\providecommand \urlprefix  [0]{URL }%
\providecommand \Eprint [0]{\href }%
\providecommand \doibase [0]{http://dx.doi.org/}%
\providecommand \selectlanguage [0]{\@gobble}%
\providecommand \bibinfo  [0]{\@secondoftwo}%
\providecommand \bibfield  [0]{\@secondoftwo}%
\providecommand \translation [1]{[#1]}%
\providecommand \BibitemOpen [0]{}%
\providecommand \bibitemStop [0]{}%
\providecommand \bibitemNoStop [0]{.\EOS\space}%
\providecommand \EOS [0]{\spacefactor3000\relax}%
\providecommand \BibitemShut  [1]{\csname bibitem#1\endcsname}%
\let\auto@bib@innerbib\@empty
%</preamble>
\bibitem [{\citenamefont {Bednorz}\ and\ \citenamefont
  {M{\"u}ller}(1986)}]{Bednorz1986}%
  \BibitemOpen
  \bibfield  {author} {\bibinfo {author} {\bibfnamefont {J.~G.}\ \bibnamefont
  {Bednorz}}\ and\ \bibinfo {author} {\bibfnamefont {K.~A.}\ \bibnamefont
  {M{\"u}ller}},\ }\href {\doibase 10.1007/BF01303701} {\bibfield  {journal}
  {\bibinfo  {journal} {Z. Phys., B, Condens. matter}\ }\textbf {\bibinfo
  {volume} {64}},\ \bibinfo {pages} {189} (\bibinfo {year} {1986})}\BibitemShut
  {NoStop}%
\bibitem [{\citenamefont {Anderson}(1987)}]{Anderson1987}%
  \BibitemOpen
  \bibfield  {author} {\bibinfo {author} {\bibfnamefont {P.~W.}\ \bibnamefont
  {Anderson}},\ }\href {\doibase 10.1126/science.235.4793.1196} {\bibfield
  {journal} {\bibinfo  {journal} {Science}\ }\textbf {\bibinfo {volume}
  {235}},\ \bibinfo {pages} {1196} (\bibinfo {year} {1987})}\BibitemShut
  {NoStop}%
\bibitem [{\citenamefont {Dagotto}(1994)}]{Dagotto1994}%
  \BibitemOpen
  \bibfield  {author} {\bibinfo {author} {\bibfnamefont {E.}~\bibnamefont
  {Dagotto}},\ }\href {\doibase 10.1103/RevModPhys.66.763} {\bibfield
  {journal} {\bibinfo  {journal} {Rev. Mod. Phys.}\ }\textbf {\bibinfo {volume}
  {66}},\ \bibinfo {pages} {763} (\bibinfo {year} {1994})}\BibitemShut
  {NoStop}%
\bibitem [{\citenamefont {Lee}\ \emph {et~al.}(2006)\citenamefont {Lee},
  \citenamefont {Nagaosa},\ and\ \citenamefont {Wen}}]{Lee2006}%
  \BibitemOpen
  \bibfield  {author} {\bibinfo {author} {\bibfnamefont {P.~A.}\ \bibnamefont
  {Lee}}, \bibinfo {author} {\bibfnamefont {N.}~\bibnamefont {Nagaosa}}, \ and\
  \bibinfo {author} {\bibfnamefont {X.-G.}\ \bibnamefont {Wen}},\ }\href
  {\doibase 10.1103/RevModPhys.78.17} {\bibfield  {journal} {\bibinfo
  {journal} {Rev. Mod. Phys.}\ }\textbf {\bibinfo {volume} {78}},\ \bibinfo
  {pages} {17} (\bibinfo {year} {2006})}\BibitemShut {NoStop}%
\bibitem [{\citenamefont {Scalapino}(2012)}]{Scalapino2012}%
  \BibitemOpen
  \bibfield  {author} {\bibinfo {author} {\bibfnamefont {D.~J.}\ \bibnamefont
  {Scalapino}},\ }\href {\doibase 10.1103/RevModPhys.84.1383} {\bibfield
  {journal} {\bibinfo  {journal} {Rev. Mod. Phys.}\ }\textbf {\bibinfo {volume}
  {84}},\ \bibinfo {pages} {1383} (\bibinfo {year} {2012})}\BibitemShut
  {NoStop}%
\bibitem [{\citenamefont {Demler}\ \emph {et~al.}(2004)\citenamefont {Demler},
  \citenamefont {Hanke},\ and\ \citenamefont {Zhang}}]{Demler2004}%
  \BibitemOpen
  \bibfield  {author} {\bibinfo {author} {\bibfnamefont {E.}~\bibnamefont
  {Demler}}, \bibinfo {author} {\bibfnamefont {W.}~\bibnamefont {Hanke}}, \
  and\ \bibinfo {author} {\bibfnamefont {S.-C.}\ \bibnamefont {Zhang}},\ }\href
  {\doibase 10.1103/RevModPhys.76.909} {\bibfield  {journal} {\bibinfo
  {journal} {Rev. Mod. Phys.}\ }\textbf {\bibinfo {volume} {76}},\ \bibinfo
  {pages} {909} (\bibinfo {year} {2004})}\BibitemShut {NoStop}%
\bibitem [{\citenamefont {Berg}\ \emph {et~al.}(2007)\citenamefont {Berg},
  \citenamefont {Fradkin}, \citenamefont {Kim}, \citenamefont {Kivelson},
  \citenamefont {Oganesyan}, \citenamefont {Tranquada},\ and\ \citenamefont
  {Zhang}}]{Berg2007}%
  \BibitemOpen
  \bibfield  {author} {\bibinfo {author} {\bibfnamefont {E.}~\bibnamefont
  {Berg}}, \bibinfo {author} {\bibfnamefont {E.}~\bibnamefont {Fradkin}},
  \bibinfo {author} {\bibfnamefont {E.-A.}\ \bibnamefont {Kim}}, \bibinfo
  {author} {\bibfnamefont {S.~A.}\ \bibnamefont {Kivelson}}, \bibinfo {author}
  {\bibfnamefont {V.}~\bibnamefont {Oganesyan}}, \bibinfo {author}
  {\bibfnamefont {J.~M.}\ \bibnamefont {Tranquada}}, \ and\ \bibinfo {author}
  {\bibfnamefont {S.~C.}\ \bibnamefont {Zhang}},\ }\href {\doibase
  10.1103/PhysRevLett.99.127003} {\bibfield  {journal} {\bibinfo  {journal}
  {Phys. Rev. Lett.}\ }\textbf {\bibinfo {volume} {99}},\ \bibinfo {pages}
  {127003} (\bibinfo {year} {2007})}\BibitemShut {NoStop}%
\bibitem [{\citenamefont {Fradkin}\ \emph {et~al.}(2015)\citenamefont
  {Fradkin}, \citenamefont {Kivelson},\ and\ \citenamefont
  {Tranquada}}]{Fradkin2015}%
  \BibitemOpen
  \bibfield  {author} {\bibinfo {author} {\bibfnamefont {E.}~\bibnamefont
  {Fradkin}}, \bibinfo {author} {\bibfnamefont {S.~A.}\ \bibnamefont
  {Kivelson}}, \ and\ \bibinfo {author} {\bibfnamefont {J.~M.}\ \bibnamefont
  {Tranquada}},\ }\href {\doibase 10.1103/RevModPhys.87.457} {\bibfield
  {journal} {\bibinfo  {journal} {Rev. Mod. Phys.}\ }\textbf {\bibinfo {volume}
  {87}},\ \bibinfo {pages} {457} (\bibinfo {year} {2015})}\BibitemShut
  {NoStop}%
\bibitem [{\citenamefont {Keimer}\ \emph {et~al.}(2015)\citenamefont {Keimer},
  \citenamefont {Kivelson}, \citenamefont {Norman}, \citenamefont {Uchida},\
  and\ \citenamefont {Zaanen}}]{Keimer2015}%
  \BibitemOpen
  \bibfield  {author} {\bibinfo {author} {\bibfnamefont {B.}~\bibnamefont
  {Keimer}}, \bibinfo {author} {\bibfnamefont {S.~A.}\ \bibnamefont
  {Kivelson}}, \bibinfo {author} {\bibfnamefont {M.~R.}\ \bibnamefont
  {Norman}}, \bibinfo {author} {\bibfnamefont {S.}~\bibnamefont {Uchida}}, \
  and\ \bibinfo {author} {\bibfnamefont {J.}~\bibnamefont {Zaanen}},\ }\href
  {http://dx.doi.org/10.1038/nature14165} {\bibfield  {journal} {\bibinfo
  {journal} {Nature}\ }\textbf {\bibinfo {volume} {518}},\ \bibinfo {pages}
  {179 EP } (\bibinfo {year} {2015})}\BibitemShut {NoStop}%
\bibitem [{\citenamefont {Kosterlitz}\ and\ \citenamefont
  {Thouless}(1973)}]{Kosterlitz1973}%
  \BibitemOpen
  \bibfield  {author} {\bibinfo {author} {\bibfnamefont {J.~M.}\ \bibnamefont
  {Kosterlitz}}\ and\ \bibinfo {author} {\bibfnamefont {D.~J.}\ \bibnamefont
  {Thouless}},\ }\href {http://stacks.iop.org/0022-3719/6/i=7/a=010} {\bibfield
   {journal} {\bibinfo  {journal} {Journal of Physics C: Solid State Physics}\
  }\textbf {\bibinfo {volume} {6}},\ \bibinfo {pages} {1181} (\bibinfo {year}
  {1973})}\BibitemShut {NoStop}%
\bibitem [{\citenamefont {Uemura}\ \emph {et~al.}(1989)\citenamefont {Uemura},
  \citenamefont {Luke}, \citenamefont {Sternlieb}, \citenamefont {Brewer},
  \citenamefont {Carolan}, \citenamefont {Hardy}, \citenamefont {Kadono},
  \citenamefont {Kempton}, \citenamefont {Kiefl}, \citenamefont {Kreitzman},
  \citenamefont {Mulhern}, \citenamefont {Riseman}, \citenamefont {Williams},
  \citenamefont {Yang}, \citenamefont {Uchida}, \citenamefont {Takagi},
  \citenamefont {Gopalakrishnan}, \citenamefont {Sleight}, \citenamefont
  {Subramanian}, \citenamefont {Chien}, \citenamefont {Cieplak}, \citenamefont
  {Xiao}, \citenamefont {Lee}, \citenamefont {Statt}, \citenamefont {Stronach},
  \citenamefont {Kossler},\ and\ \citenamefont {Yu}}]{Uemura1989}%
  \BibitemOpen
  \bibfield  {author} {\bibinfo {author} {\bibfnamefont {Y.~J.}\ \bibnamefont
  {Uemura}}, \bibinfo {author} {\bibfnamefont {G.~M.}\ \bibnamefont {Luke}},
  \bibinfo {author} {\bibfnamefont {B.~J.}\ \bibnamefont {Sternlieb}}, \bibinfo
  {author} {\bibfnamefont {J.~H.}\ \bibnamefont {Brewer}}, \bibinfo {author}
  {\bibfnamefont {J.~F.}\ \bibnamefont {Carolan}}, \bibinfo {author}
  {\bibfnamefont {W.~N.}\ \bibnamefont {Hardy}}, \bibinfo {author}
  {\bibfnamefont {R.}~\bibnamefont {Kadono}}, \bibinfo {author} {\bibfnamefont
  {J.~R.}\ \bibnamefont {Kempton}}, \bibinfo {author} {\bibfnamefont {R.~F.}\
  \bibnamefont {Kiefl}}, \bibinfo {author} {\bibfnamefont {S.~R.}\ \bibnamefont
  {Kreitzman}}, \bibinfo {author} {\bibfnamefont {P.}~\bibnamefont {Mulhern}},
  \bibinfo {author} {\bibfnamefont {T.~M.}\ \bibnamefont {Riseman}}, \bibinfo
  {author} {\bibfnamefont {D.~L.}\ \bibnamefont {Williams}}, \bibinfo {author}
  {\bibfnamefont {B.~X.}\ \bibnamefont {Yang}}, \bibinfo {author}
  {\bibfnamefont {S.}~\bibnamefont {Uchida}}, \bibinfo {author} {\bibfnamefont
  {H.}~\bibnamefont {Takagi}}, \bibinfo {author} {\bibfnamefont
  {J.}~\bibnamefont {Gopalakrishnan}}, \bibinfo {author} {\bibfnamefont
  {A.~W.}\ \bibnamefont {Sleight}}, \bibinfo {author} {\bibfnamefont {M.~A.}\
  \bibnamefont {Subramanian}}, \bibinfo {author} {\bibfnamefont {C.~L.}\
  \bibnamefont {Chien}}, \bibinfo {author} {\bibfnamefont {M.~Z.}\ \bibnamefont
  {Cieplak}}, \bibinfo {author} {\bibfnamefont {G.}~\bibnamefont {Xiao}},
  \bibinfo {author} {\bibfnamefont {V.~Y.}\ \bibnamefont {Lee}}, \bibinfo
  {author} {\bibfnamefont {B.~W.}\ \bibnamefont {Statt}}, \bibinfo {author}
  {\bibfnamefont {C.~E.}\ \bibnamefont {Stronach}}, \bibinfo {author}
  {\bibfnamefont {W.~J.}\ \bibnamefont {Kossler}}, \ and\ \bibinfo {author}
  {\bibfnamefont {X.~H.}\ \bibnamefont {Yu}},\ }\href {\doibase
  10.1103/PhysRevLett.62.2317} {\bibfield  {journal} {\bibinfo  {journal}
  {Phys. Rev. Lett.}\ }\textbf {\bibinfo {volume} {62}},\ \bibinfo {pages}
  {2317} (\bibinfo {year} {1989})}\BibitemShut {NoStop}%
\bibitem [{\citenamefont {Emery}\ and\ \citenamefont
  {Kivelson}(1995)}]{Emery1995}%
  \BibitemOpen
  \bibfield  {author} {\bibinfo {author} {\bibfnamefont {V.~J.}\ \bibnamefont
  {Emery}}\ and\ \bibinfo {author} {\bibfnamefont {S.~A.}\ \bibnamefont
  {Kivelson}},\ }\href {http://dx.doi.org/10.1038/374434a0} {\bibfield
  {journal} {\bibinfo  {journal} {Nature}\ }\textbf {\bibinfo {volume} {374}},\
  \bibinfo {pages} {434 EP } (\bibinfo {year} {1995})}\BibitemShut {NoStop}%
\bibitem [{\citenamefont {Alvarez}\ and\ \citenamefont
  {Balseiro}(1996)}]{Alvarez1996}%
  \BibitemOpen
  \bibfield  {author} {\bibinfo {author} {\bibfnamefont {J.~V.}\ \bibnamefont
  {Alvarez}}\ and\ \bibinfo {author} {\bibfnamefont {C.}~\bibnamefont
  {Balseiro}},\ }\href {\doibase https://doi.org/10.1016/0038-1098(96)00047-6}
  {\bibfield  {journal} {\bibinfo  {journal} {Solid State Communications}\
  }\textbf {\bibinfo {volume} {98}},\ \bibinfo {pages} {313 } (\bibinfo {year}
  {1996})}\BibitemShut {NoStop}%
\bibitem [{\citenamefont {Kwon}\ \emph {et~al.}(2001)\citenamefont {Kwon},
  \citenamefont {Dorsey},\ and\ \citenamefont {Hirschfeld}}]{Kwon2001}%
  \BibitemOpen
  \bibfield  {author} {\bibinfo {author} {\bibfnamefont {H.-J.}\ \bibnamefont
  {Kwon}}, \bibinfo {author} {\bibfnamefont {A.~T.}\ \bibnamefont {Dorsey}}, \
  and\ \bibinfo {author} {\bibfnamefont {P.~J.}\ \bibnamefont {Hirschfeld}},\
  }\href {\doibase 10.1103/PhysRevLett.86.3875} {\bibfield  {journal} {\bibinfo
   {journal} {Phys. Rev. Lett.}\ }\textbf {\bibinfo {volume} {86}},\ \bibinfo
  {pages} {3875} (\bibinfo {year} {2001})}\BibitemShut {NoStop}%
\bibitem [{\citenamefont {Sharapov}\ \emph {et~al.}(2001)\citenamefont
  {Sharapov}, \citenamefont {Beck},\ and\ \citenamefont
  {Loktev}}]{Sharapov2001}%
  \BibitemOpen
  \bibfield  {author} {\bibinfo {author} {\bibfnamefont {S.~G.}\ \bibnamefont
  {Sharapov}}, \bibinfo {author} {\bibfnamefont {H.}~\bibnamefont {Beck}}, \
  and\ \bibinfo {author} {\bibfnamefont {V.~M.}\ \bibnamefont {Loktev}},\
  }\href {\doibase 10.1103/PhysRevB.64.134519} {\bibfield  {journal} {\bibinfo
  {journal} {Phys. Rev. B}\ }\textbf {\bibinfo {volume} {64}},\ \bibinfo
  {pages} {134519} (\bibinfo {year} {2001})}\BibitemShut {NoStop}%
\bibitem [{\citenamefont {Benfatto}\ \emph {et~al.}(2004)\citenamefont
  {Benfatto}, \citenamefont {Toschi},\ and\ \citenamefont
  {Caprara}}]{Benfatto2004}%
  \BibitemOpen
  \bibfield  {author} {\bibinfo {author} {\bibfnamefont {L.}~\bibnamefont
  {Benfatto}}, \bibinfo {author} {\bibfnamefont {A.}~\bibnamefont {Toschi}}, \
  and\ \bibinfo {author} {\bibfnamefont {S.}~\bibnamefont {Caprara}},\ }\href
  {\doibase 10.1103/PhysRevB.69.184510} {\bibfield  {journal} {\bibinfo
  {journal} {Phys. Rev. B}\ }\textbf {\bibinfo {volume} {69}},\ \bibinfo
  {pages} {184510} (\bibinfo {year} {2004})}\BibitemShut {NoStop}%
\bibitem [{\citenamefont {Imada}\ \emph {et~al.}(1998)\citenamefont {Imada},
  \citenamefont {Fujimori},\ and\ \citenamefont {Tokura}}]{Imada1998}%
  \BibitemOpen
  \bibfield  {author} {\bibinfo {author} {\bibfnamefont {M.}~\bibnamefont
  {Imada}}, \bibinfo {author} {\bibfnamefont {A.}~\bibnamefont {Fujimori}}, \
  and\ \bibinfo {author} {\bibfnamefont {Y.}~\bibnamefont {Tokura}},\ }\href
  {\doibase 10.1103/RevModPhys.70.1039} {\bibfield  {journal} {\bibinfo
  {journal} {Rev. Mod. Phys.}\ }\textbf {\bibinfo {volume} {70}},\ \bibinfo
  {pages} {1039} (\bibinfo {year} {1998})}\BibitemShut {NoStop}%
\bibitem [{\citenamefont {Lichtenstein}\ and\ \citenamefont
  {Katsnelson}(2000)}]{Lichtenstein2000}%
  \BibitemOpen
  \bibfield  {author} {\bibinfo {author} {\bibfnamefont {A.~I.}\ \bibnamefont
  {Lichtenstein}}\ and\ \bibinfo {author} {\bibfnamefont {M.~I.}\ \bibnamefont
  {Katsnelson}},\ }\href {\doibase 10.1103/PhysRevB.62.R9283} {\bibfield
  {journal} {\bibinfo  {journal} {Phys. Rev. B}\ }\textbf {\bibinfo {volume}
  {62}},\ \bibinfo {pages} {R9283} (\bibinfo {year} {2000})}\BibitemShut
  {NoStop}%
\bibitem [{\citenamefont {Kotliar}\ \emph {et~al.}(2001)\citenamefont
  {Kotliar}, \citenamefont {Savrasov}, \citenamefont {P\'alsson},\ and\
  \citenamefont {Biroli}}]{Kotliar2001}%
  \BibitemOpen
  \bibfield  {author} {\bibinfo {author} {\bibfnamefont {G.}~\bibnamefont
  {Kotliar}}, \bibinfo {author} {\bibfnamefont {S.~Y.}\ \bibnamefont
  {Savrasov}}, \bibinfo {author} {\bibfnamefont {G.}~\bibnamefont {P\'alsson}},
  \ and\ \bibinfo {author} {\bibfnamefont {G.}~\bibnamefont {Biroli}},\ }\href
  {\doibase 10.1103/PhysRevLett.87.186401} {\bibfield  {journal} {\bibinfo
  {journal} {Phys. Rev. Lett.}\ }\textbf {\bibinfo {volume} {87}},\ \bibinfo
  {pages} {186401} (\bibinfo {year} {2001})}\BibitemShut {NoStop}%
\bibitem [{\citenamefont {Maier}\ \emph {et~al.}(2005)\citenamefont {Maier},
  \citenamefont {Jarrell}, \citenamefont {Pruschke},\ and\ \citenamefont
  {Hettler}}]{Maier2005}%
  \BibitemOpen
  \bibfield  {author} {\bibinfo {author} {\bibfnamefont {T.~A.}\ \bibnamefont
  {Maier}}, \bibinfo {author} {\bibfnamefont {M.}~\bibnamefont {Jarrell}},
  \bibinfo {author} {\bibfnamefont {T.}~\bibnamefont {Pruschke}}, \ and\
  \bibinfo {author} {\bibfnamefont {M.~H.}\ \bibnamefont {Hettler}},\ }\href
  {\doibase 10.1103/RevModPhys.77.1027} {\bibfield  {journal} {\bibinfo
  {journal} {Rev. Mod. Phys.}\ }\textbf {\bibinfo {volume} {77}},\ \bibinfo
  {pages} {1027} (\bibinfo {year} {2005})}\BibitemShut {NoStop}%
\bibitem [{\citenamefont {Haule}\ and\ \citenamefont
  {Kotliar}(2007)}]{Haule2007}%
  \BibitemOpen
  \bibfield  {author} {\bibinfo {author} {\bibfnamefont {K.}~\bibnamefont
  {Haule}}\ and\ \bibinfo {author} {\bibfnamefont {G.}~\bibnamefont
  {Kotliar}},\ }\href {\doibase 10.1103/PhysRevB.76.104509} {\bibfield
  {journal} {\bibinfo  {journal} {Phys. Rev. B}\ }\textbf {\bibinfo {volume}
  {76}},\ \bibinfo {pages} {104509} (\bibinfo {year} {2007})}\BibitemShut
  {NoStop}%
\bibitem [{\citenamefont {Civelli}\ \emph {et~al.}(2008)\citenamefont
  {Civelli}, \citenamefont {Capone}, \citenamefont {Georges}, \citenamefont
  {Haule}, \citenamefont {Parcollet}, \citenamefont {Stanescu},\ and\
  \citenamefont {Kotliar}}]{Civelli2008}%
  \BibitemOpen
  \bibfield  {author} {\bibinfo {author} {\bibfnamefont {M.}~\bibnamefont
  {Civelli}}, \bibinfo {author} {\bibfnamefont {M.}~\bibnamefont {Capone}},
  \bibinfo {author} {\bibfnamefont {A.}~\bibnamefont {Georges}}, \bibinfo
  {author} {\bibfnamefont {K.}~\bibnamefont {Haule}}, \bibinfo {author}
  {\bibfnamefont {O.}~\bibnamefont {Parcollet}}, \bibinfo {author}
  {\bibfnamefont {T.~D.}\ \bibnamefont {Stanescu}}, \ and\ \bibinfo {author}
  {\bibfnamefont {G.}~\bibnamefont {Kotliar}},\ }\href {\doibase
  10.1103/PhysRevLett.100.046402} {\bibfield  {journal} {\bibinfo  {journal}
  {Phys. Rev. Lett.}\ }\textbf {\bibinfo {volume} {100}},\ \bibinfo {pages}
  {046402} (\bibinfo {year} {2008})}\BibitemShut {NoStop}%
\bibitem [{\citenamefont {Ferrero}\ \emph {et~al.}(2009)\citenamefont
  {Ferrero}, \citenamefont {Cornaglia}, \citenamefont {De~Leo}, \citenamefont
  {Parcollet}, \citenamefont {Kotliar},\ and\ \citenamefont
  {Georges}}]{Ferrero2009}%
  \BibitemOpen
  \bibfield  {author} {\bibinfo {author} {\bibfnamefont {M.}~\bibnamefont
  {Ferrero}}, \bibinfo {author} {\bibfnamefont {P.~S.}\ \bibnamefont
  {Cornaglia}}, \bibinfo {author} {\bibfnamefont {L.}~\bibnamefont {De~Leo}},
  \bibinfo {author} {\bibfnamefont {O.}~\bibnamefont {Parcollet}}, \bibinfo
  {author} {\bibfnamefont {G.}~\bibnamefont {Kotliar}}, \ and\ \bibinfo
  {author} {\bibfnamefont {A.}~\bibnamefont {Georges}},\ }\href {\doibase
  10.1103/PhysRevB.80.064501} {\bibfield  {journal} {\bibinfo  {journal} {Phys.
  Rev. B}\ }\textbf {\bibinfo {volume} {80}},\ \bibinfo {pages} {064501}
  (\bibinfo {year} {2009})}\BibitemShut {NoStop}%
\bibitem [{\citenamefont {Sordi}\ \emph {et~al.}(2012)\citenamefont {Sordi},
  \citenamefont {S\'emon}, \citenamefont {Haule},\ and\ \citenamefont
  {Tremblay}}]{Sordi2012}%
  \BibitemOpen
  \bibfield  {author} {\bibinfo {author} {\bibfnamefont {G.}~\bibnamefont
  {Sordi}}, \bibinfo {author} {\bibfnamefont {P.}~\bibnamefont {S\'emon}},
  \bibinfo {author} {\bibfnamefont {K.}~\bibnamefont {Haule}}, \ and\ \bibinfo
  {author} {\bibfnamefont {A.-M.~S.}\ \bibnamefont {Tremblay}},\ }\href
  {\doibase 10.1103/PhysRevLett.108.216401} {\bibfield  {journal} {\bibinfo
  {journal} {Phys. Rev. Lett.}\ }\textbf {\bibinfo {volume} {108}},\ \bibinfo
  {pages} {216401} (\bibinfo {year} {2012})}\BibitemShut {NoStop}%
\bibitem [{\citenamefont {Gull}\ \emph {et~al.}(2013)\citenamefont {Gull},
  \citenamefont {Parcollet},\ and\ \citenamefont {Millis}}]{Gull2013}%
  \BibitemOpen
  \bibfield  {author} {\bibinfo {author} {\bibfnamefont {E.}~\bibnamefont
  {Gull}}, \bibinfo {author} {\bibfnamefont {O.}~\bibnamefont {Parcollet}}, \
  and\ \bibinfo {author} {\bibfnamefont {A.~J.}\ \bibnamefont {Millis}},\
  }\href {\doibase 10.1103/PhysRevLett.110.216405} {\bibfield  {journal}
  {\bibinfo  {journal} {Phys. Rev. Lett.}\ }\textbf {\bibinfo {volume} {110}},\
  \bibinfo {pages} {216405} (\bibinfo {year} {2013})}\BibitemShut {NoStop}%
\bibitem [{\citenamefont {H\'ebert}\ \emph {et~al.}(2015)\citenamefont
  {H\'ebert}, \citenamefont {S\'emon},\ and\ \citenamefont
  {Tremblay}}]{Hebert2015}%
  \BibitemOpen
  \bibfield  {author} {\bibinfo {author} {\bibfnamefont {C.-D.}\ \bibnamefont
  {H\'ebert}}, \bibinfo {author} {\bibfnamefont {P.}~\bibnamefont {S\'emon}}, \
  and\ \bibinfo {author} {\bibfnamefont {A.-M.~S.}\ \bibnamefont {Tremblay}},\
  }\href {\doibase 10.1103/PhysRevB.92.195112} {\bibfield  {journal} {\bibinfo
  {journal} {Phys. Rev. B}\ }\textbf {\bibinfo {volume} {92}},\ \bibinfo
  {pages} {195112} (\bibinfo {year} {2015})}\BibitemShut {NoStop}%
\bibitem [{\citenamefont {Fratino}\ \emph {et~al.}(2016)\citenamefont
  {Fratino}, \citenamefont {S{\'e}mon}, \citenamefont {Sordi},\ and\
  \citenamefont {Tremblay}}]{Fratino2016}%
  \BibitemOpen
  \bibfield  {author} {\bibinfo {author} {\bibfnamefont {L.}~\bibnamefont
  {Fratino}}, \bibinfo {author} {\bibfnamefont {P.}~\bibnamefont {S{\'e}mon}},
  \bibinfo {author} {\bibfnamefont {G.}~\bibnamefont {Sordi}}, \ and\ \bibinfo
  {author} {\bibfnamefont {A.-M.~S.}\ \bibnamefont {Tremblay}},\ }\href
  {http://dx.doi.org/10.1038/srep22715} {\bibfield  {journal} {\bibinfo
  {journal} {Scientific Reports}\ }\textbf {\bibinfo {volume} {6}},\ \bibinfo
  {pages} {22715 EP } (\bibinfo {year} {2016})}\BibitemShut {NoStop}%
\bibitem [{\citenamefont {Harland}\ \emph {et~al.}(2016)\citenamefont
  {Harland}, \citenamefont {Katsnelson},\ and\ \citenamefont
  {Lichtenstein}}]{Harland2016}%
  \BibitemOpen
  \bibfield  {author} {\bibinfo {author} {\bibfnamefont {M.}~\bibnamefont
  {Harland}}, \bibinfo {author} {\bibfnamefont {M.~I.}\ \bibnamefont
  {Katsnelson}}, \ and\ \bibinfo {author} {\bibfnamefont {A.~I.}\ \bibnamefont
  {Lichtenstein}},\ }\href {\doibase 10.1103/PhysRevB.94.125133} {\bibfield
  {journal} {\bibinfo  {journal} {Phys. Rev. B}\ }\textbf {\bibinfo {volume}
  {94}},\ \bibinfo {pages} {125133} (\bibinfo {year} {2016})}\BibitemShut
  {NoStop}%
\bibitem [{\citenamefont {Bragan\ifmmode~\mbox{\c{c}}\else \c{c}\fi{}a}\ \emph
  {et~al.}(2018)\citenamefont {Bragan\ifmmode~\mbox{\c{c}}\else \c{c}\fi{}a},
  \citenamefont {Sakai}, \citenamefont {Aguiar},\ and\ \citenamefont
  {Civelli}}]{Braganca2018}%
  \BibitemOpen
  \bibfield  {author} {\bibinfo {author} {\bibfnamefont {H.}~\bibnamefont
  {Bragan\ifmmode~\mbox{\c{c}}\else \c{c}\fi{}a}}, \bibinfo {author}
  {\bibfnamefont {S.}~\bibnamefont {Sakai}}, \bibinfo {author} {\bibfnamefont
  {M.~C.~O.}\ \bibnamefont {Aguiar}}, \ and\ \bibinfo {author} {\bibfnamefont
  {M.}~\bibnamefont {Civelli}},\ }\href {\doibase
  10.1103/PhysRevLett.120.067002} {\bibfield  {journal} {\bibinfo  {journal}
  {Phys. Rev. Lett.}\ }\textbf {\bibinfo {volume} {120}},\ \bibinfo {pages}
  {067002} (\bibinfo {year} {2018})}\BibitemShut {NoStop}%
\bibitem [{\citenamefont {Wu}\ \emph {et~al.}(2018)\citenamefont {Wu},
  \citenamefont {Scheurer}, \citenamefont {Chatterjee}, \citenamefont
  {Sachdev}, \citenamefont {Georges},\ and\ \citenamefont {Ferrero}}]{Wu2018}%
  \BibitemOpen
  \bibfield  {author} {\bibinfo {author} {\bibfnamefont {W.}~\bibnamefont
  {Wu}}, \bibinfo {author} {\bibfnamefont {M.~S.}\ \bibnamefont {Scheurer}},
  \bibinfo {author} {\bibfnamefont {S.}~\bibnamefont {Chatterjee}}, \bibinfo
  {author} {\bibfnamefont {S.}~\bibnamefont {Sachdev}}, \bibinfo {author}
  {\bibfnamefont {A.}~\bibnamefont {Georges}}, \ and\ \bibinfo {author}
  {\bibfnamefont {M.}~\bibnamefont {Ferrero}},\ }\href {\doibase
  10.1103/PhysRevX.8.021048} {\bibfield  {journal} {\bibinfo  {journal} {Phys.
  Rev. X}\ }\textbf {\bibinfo {volume} {8}},\ \bibinfo {pages} {021048}
  (\bibinfo {year} {2018})}\BibitemShut {NoStop}%
\bibitem [{\citenamefont {{Jiang}}\ and\ \citenamefont
  {{Devereaux}}(2018)}]{Jiang2018}%
  \BibitemOpen
  \bibfield  {author} {\bibinfo {author} {\bibfnamefont {H.-C.}\ \bibnamefont
  {{Jiang}}}\ and\ \bibinfo {author} {\bibfnamefont {T.~P.}\ \bibnamefont
  {{Devereaux}}},\ }\href@noop {} {\bibfield  {journal} {\bibinfo  {journal}
  {arXiv e-prints}\ ,\ \bibinfo {eid} {arXiv:1806.01465}} (\bibinfo {year}
  {2018})}\BibitemShut {NoStop}%
\bibitem [{\citenamefont {Jiang}\ \emph {et~al.}(2018)\citenamefont {Jiang},
  \citenamefont {Weng},\ and\ \citenamefont {Kivelson}}]{Jiang2018a}%
  \BibitemOpen
  \bibfield  {author} {\bibinfo {author} {\bibfnamefont {H.-C.}\ \bibnamefont
  {Jiang}}, \bibinfo {author} {\bibfnamefont {Z.-Y.}\ \bibnamefont {Weng}}, \
  and\ \bibinfo {author} {\bibfnamefont {S.~A.}\ \bibnamefont {Kivelson}},\
  }\href {\doibase 10.1103/PhysRevB.98.140505} {\bibfield  {journal} {\bibinfo
  {journal} {Phys. Rev. B}\ }\textbf {\bibinfo {volume} {98}},\ \bibinfo
  {pages} {140505} (\bibinfo {year} {2018})}\BibitemShut {NoStop}%
\bibitem [{\citenamefont {Liechtenstein}\ \emph {et~al.}(1987)\citenamefont
  {Liechtenstein}, \citenamefont {Katsnelson}, \citenamefont {Antropov},\ and\
  \citenamefont {Gubanov}}]{Liechtenstein1987}%
  \BibitemOpen
  \bibfield  {author} {\bibinfo {author} {\bibfnamefont {A.}~\bibnamefont
  {Liechtenstein}}, \bibinfo {author} {\bibfnamefont {M.}~\bibnamefont
  {Katsnelson}}, \bibinfo {author} {\bibfnamefont {V.}~\bibnamefont
  {Antropov}}, \ and\ \bibinfo {author} {\bibfnamefont {V.}~\bibnamefont
  {Gubanov}},\ }\href {\doibase https://doi.org/10.1016/0304-8853(87)90721-9}
  {\bibfield  {journal} {\bibinfo  {journal} {J. Magn. Magn. Mater.}\ }\textbf
  {\bibinfo {volume} {67}},\ \bibinfo {pages} {65 } (\bibinfo {year}
  {1987})}\BibitemShut {NoStop}%
\bibitem [{\citenamefont {Katsnelson}\ and\ \citenamefont
  {Lichtenstein}(2000)}]{Katsnelson2000}%
  \BibitemOpen
  \bibfield  {author} {\bibinfo {author} {\bibfnamefont {M.~I.}\ \bibnamefont
  {Katsnelson}}\ and\ \bibinfo {author} {\bibfnamefont {A.~I.}\ \bibnamefont
  {Lichtenstein}},\ }\href {\doibase 10.1103/PhysRevB.61.8906} {\bibfield
  {journal} {\bibinfo  {journal} {Phys. Rev. B}\ }\textbf {\bibinfo {volume}
  {61}},\ \bibinfo {pages} {8906} (\bibinfo {year} {2000})}\BibitemShut
  {NoStop}%
\bibitem [{\citenamefont {Eriksson}\ \emph {et~al.}(2017)\citenamefont
  {Eriksson}, \citenamefont {Bergman}, \citenamefont {Bergqvist},\ and\
  \citenamefont {Hellsvik}}]{Eriksson2017}%
  \BibitemOpen
  \bibfield  {author} {\bibinfo {author} {\bibfnamefont {O.}~\bibnamefont
  {Eriksson}}, \bibinfo {author} {\bibfnamefont {A.}~\bibnamefont {Bergman}},
  \bibinfo {author} {\bibfnamefont {L.}~\bibnamefont {Bergqvist}}, \ and\
  \bibinfo {author} {\bibfnamefont {J.}~\bibnamefont {Hellsvik}},\ }\href
  {http://www.oxfordscholarship.com/10.1093/oso/9780198788669.001.0001/oso-9780198788669}
  {\emph {\bibinfo {title} {Atomistic Spin Dynamics: Foundations and
  Applications}}}\ (\bibinfo  {publisher} {Oxford University Press},\ \bibinfo
  {address} {Oxford},\ \bibinfo {year} {2017})\ p.\ \bibinfo {pages}
  {272}\BibitemShut {NoStop}%
\bibitem [{\citenamefont {Hubbard}(1963)}]{Hubbard1963}%
  \BibitemOpen
  \bibfield  {author} {\bibinfo {author} {\bibfnamefont {J.}~\bibnamefont
  {Hubbard}},\ }\href {\doibase 10.1098/rspa.1963.0204} {\bibfield  {journal}
  {\bibinfo  {journal} {Proc. R. Soc. Lond. A}\ }\textbf {\bibinfo {volume}
  {276}},\ \bibinfo {pages} {238} (\bibinfo {year} {1963})}\BibitemShut
  {NoStop}%
\bibitem [{\citenamefont {Pavarini}\ \emph {et~al.}(2001)\citenamefont
  {Pavarini}, \citenamefont {Dasgupta}, \citenamefont {Saha-Dasgupta},
  \citenamefont {Jepsen},\ and\ \citenamefont {Andersen}}]{Pavarini2001}%
  \BibitemOpen
  \bibfield  {author} {\bibinfo {author} {\bibfnamefont {E.}~\bibnamefont
  {Pavarini}}, \bibinfo {author} {\bibfnamefont {I.}~\bibnamefont {Dasgupta}},
  \bibinfo {author} {\bibfnamefont {T.}~\bibnamefont {Saha-Dasgupta}}, \bibinfo
  {author} {\bibfnamefont {O.}~\bibnamefont {Jepsen}}, \ and\ \bibinfo {author}
  {\bibfnamefont {O.~K.}\ \bibnamefont {Andersen}},\ }\href {\doibase
  10.1103/PhysRevLett.87.047003} {\bibfield  {journal} {\bibinfo  {journal}
  {Phys. Rev. Lett.}\ }\textbf {\bibinfo {volume} {87}},\ \bibinfo {pages}
  {047003} (\bibinfo {year} {2001})}\BibitemShut {NoStop}%
\bibitem [{\citenamefont {Chakravarty}\ \emph {et~al.}(1993)\citenamefont
  {Chakravarty}, \citenamefont {Sudb{\o}}, \citenamefont {Anderson},\ and\
  \citenamefont {Strong}}]{Chakravarty1993}%
  \BibitemOpen
  \bibfield  {author} {\bibinfo {author} {\bibfnamefont {S.}~\bibnamefont
  {Chakravarty}}, \bibinfo {author} {\bibfnamefont {A.}~\bibnamefont
  {Sudb{\o}}}, \bibinfo {author} {\bibfnamefont {P.~W.}\ \bibnamefont
  {Anderson}}, \ and\ \bibinfo {author} {\bibfnamefont {S.}~\bibnamefont
  {Strong}},\ }\href {\doibase 10.1126/science.261.5119.337} {\bibfield
  {journal} {\bibinfo  {journal} {Science}\ }\textbf {\bibinfo {volume}
  {261}},\ \bibinfo {pages} {337} (\bibinfo {year} {1993})}\BibitemShut
  {NoStop}%
\bibitem [{\citenamefont {Andersen}\ \emph {et~al.}(1995)\citenamefont
  {Andersen}, \citenamefont {Liechtenstein}, \citenamefont {Jepsen},\ and\
  \citenamefont {Paulsen}}]{Andersen1995}%
  \BibitemOpen
  \bibfield  {author} {\bibinfo {author} {\bibfnamefont {O.}~\bibnamefont
  {Andersen}}, \bibinfo {author} {\bibfnamefont {A.}~\bibnamefont
  {Liechtenstein}}, \bibinfo {author} {\bibfnamefont {O.}~\bibnamefont
  {Jepsen}}, \ and\ \bibinfo {author} {\bibfnamefont {F.}~\bibnamefont
  {Paulsen}},\ }\href {\doibase https://doi.org/10.1016/0022-3697(95)00269-3}
  {\bibfield  {journal} {\bibinfo  {journal} {J. Phys. Chem. Solids}\ }\textbf
  {\bibinfo {volume} {56}},\ \bibinfo {pages} {1573 } (\bibinfo {year}
  {1995})}\BibitemShut {NoStop}%
\bibitem [{\citenamefont {Wu}\ \emph {et~al.}(1987)\citenamefont {Wu},
  \citenamefont {Ashburn}, \citenamefont {Torng}, \citenamefont {Hor},
  \citenamefont {Meng}, \citenamefont {Gao}, \citenamefont {Huang},
  \citenamefont {Wang},\ and\ \citenamefont {Chu}}]{Wu1987}%
  \BibitemOpen
  \bibfield  {author} {\bibinfo {author} {\bibfnamefont {M.~K.}\ \bibnamefont
  {Wu}}, \bibinfo {author} {\bibfnamefont {J.~R.}\ \bibnamefont {Ashburn}},
  \bibinfo {author} {\bibfnamefont {C.~J.}\ \bibnamefont {Torng}}, \bibinfo
  {author} {\bibfnamefont {P.~H.}\ \bibnamefont {Hor}}, \bibinfo {author}
  {\bibfnamefont {R.~L.}\ \bibnamefont {Meng}}, \bibinfo {author}
  {\bibfnamefont {L.}~\bibnamefont {Gao}}, \bibinfo {author} {\bibfnamefont
  {Z.~J.}\ \bibnamefont {Huang}}, \bibinfo {author} {\bibfnamefont {Y.~Q.}\
  \bibnamefont {Wang}}, \ and\ \bibinfo {author} {\bibfnamefont {C.~W.}\
  \bibnamefont {Chu}},\ }\href {\doibase 10.1103/PhysRevLett.58.908} {\bibfield
   {journal} {\bibinfo  {journal} {Phys. Rev. Lett.}\ }\textbf {\bibinfo
  {volume} {58}},\ \bibinfo {pages} {908} (\bibinfo {year} {1987})}\BibitemShut
  {NoStop}%
\bibitem [{\citenamefont {Cava}\ \emph {et~al.}(1987)\citenamefont {Cava},
  \citenamefont {Batlogg}, \citenamefont {van Dover}, \citenamefont {Murphy},
  \citenamefont {Sunshine}, \citenamefont {Siegrist}, \citenamefont {Remeika},
  \citenamefont {Rietman}, \citenamefont {Zahurak},\ and\ \citenamefont
  {Espinosa}}]{Cava1987}%
  \BibitemOpen
  \bibfield  {author} {\bibinfo {author} {\bibfnamefont {R.~J.}\ \bibnamefont
  {Cava}}, \bibinfo {author} {\bibfnamefont {B.}~\bibnamefont {Batlogg}},
  \bibinfo {author} {\bibfnamefont {R.~B.}\ \bibnamefont {van Dover}}, \bibinfo
  {author} {\bibfnamefont {D.~W.}\ \bibnamefont {Murphy}}, \bibinfo {author}
  {\bibfnamefont {S.}~\bibnamefont {Sunshine}}, \bibinfo {author}
  {\bibfnamefont {T.}~\bibnamefont {Siegrist}}, \bibinfo {author}
  {\bibfnamefont {J.~P.}\ \bibnamefont {Remeika}}, \bibinfo {author}
  {\bibfnamefont {E.~A.}\ \bibnamefont {Rietman}}, \bibinfo {author}
  {\bibfnamefont {S.}~\bibnamefont {Zahurak}}, \ and\ \bibinfo {author}
  {\bibfnamefont {G.~P.}\ \bibnamefont {Espinosa}},\ }\href {\doibase
  10.1103/PhysRevLett.58.1676} {\bibfield  {journal} {\bibinfo  {journal}
  {Phys. Rev. Lett.}\ }\textbf {\bibinfo {volume} {58}},\ \bibinfo {pages}
  {1676} (\bibinfo {year} {1987})}\BibitemShut {NoStop}%
\bibitem [{\citenamefont {Parcollet}\ \emph {et~al.}(2015)\citenamefont
  {Parcollet}, \citenamefont {Ferrero}, \citenamefont {Ayral}, \citenamefont
  {Hafermann}, \citenamefont {Krivenko}, \citenamefont {Messio},\ and\
  \citenamefont {Seth}}]{Parcollet2015}%
  \BibitemOpen
  \bibfield  {author} {\bibinfo {author} {\bibfnamefont {O.}~\bibnamefont
  {Parcollet}}, \bibinfo {author} {\bibfnamefont {M.}~\bibnamefont {Ferrero}},
  \bibinfo {author} {\bibfnamefont {T.}~\bibnamefont {Ayral}}, \bibinfo
  {author} {\bibfnamefont {H.}~\bibnamefont {Hafermann}}, \bibinfo {author}
  {\bibfnamefont {I.}~\bibnamefont {Krivenko}}, \bibinfo {author}
  {\bibfnamefont {L.}~\bibnamefont {Messio}}, \ and\ \bibinfo {author}
  {\bibfnamefont {P.}~\bibnamefont {Seth}},\ }\href {\doibase
  http://dx.doi.org/10.1016/j.cpc.2015.04.023} {\bibfield  {journal} {\bibinfo
  {journal} {Comput. Phys. Commun.}\ }\textbf {\bibinfo {volume} {196}},\
  \bibinfo {pages} {398 } (\bibinfo {year} {2015})}\BibitemShut {NoStop}%
\bibitem [{\citenamefont {Seth}\ \emph {et~al.}(2016)\citenamefont {Seth},
  \citenamefont {Krivenko}, \citenamefont {Ferrero},\ and\ \citenamefont
  {Parcollet}}]{Seth2016}%
  \BibitemOpen
  \bibfield  {author} {\bibinfo {author} {\bibfnamefont {P.}~\bibnamefont
  {Seth}}, \bibinfo {author} {\bibfnamefont {I.}~\bibnamefont {Krivenko}},
  \bibinfo {author} {\bibfnamefont {M.}~\bibnamefont {Ferrero}}, \ and\
  \bibinfo {author} {\bibfnamefont {O.}~\bibnamefont {Parcollet}},\ }\href
  {\doibase http://dx.doi.org/10.1016/j.cpc.2015.10.023} {\bibfield  {journal}
  {\bibinfo  {journal} {Comput. Phys. Commun.}\ }\textbf {\bibinfo {volume}
  {200}},\ \bibinfo {pages} {274 } (\bibinfo {year} {2016})}\BibitemShut
  {NoStop}%
\bibitem [{\citenamefont {Gull}\ \emph {et~al.}(2011)\citenamefont {Gull},
  \citenamefont {Millis}, \citenamefont {Lichtenstein}, \citenamefont
  {Rubtsov}, \citenamefont {Troyer},\ and\ \citenamefont {Werner}}]{Gull2011}%
  \BibitemOpen
  \bibfield  {author} {\bibinfo {author} {\bibfnamefont {E.}~\bibnamefont
  {Gull}}, \bibinfo {author} {\bibfnamefont {A.~J.}\ \bibnamefont {Millis}},
  \bibinfo {author} {\bibfnamefont {A.~I.}\ \bibnamefont {Lichtenstein}},
  \bibinfo {author} {\bibfnamefont {A.~N.}\ \bibnamefont {Rubtsov}}, \bibinfo
  {author} {\bibfnamefont {M.}~\bibnamefont {Troyer}}, \ and\ \bibinfo {author}
  {\bibfnamefont {P.}~\bibnamefont {Werner}},\ }\href {\doibase
  10.1103/RevModPhys.83.349} {\bibfield  {journal} {\bibinfo  {journal} {Rev.
  Mod. Phys.}\ }\textbf {\bibinfo {volume} {83}},\ \bibinfo {pages} {349}
  (\bibinfo {year} {2011})}\BibitemShut {NoStop}%
\bibitem [{\citenamefont {Luttinger}\ and\ \citenamefont
  {Ward}(1960)}]{Luttinger1960}%
  \BibitemOpen
  \bibfield  {author} {\bibinfo {author} {\bibfnamefont {J.~M.}\ \bibnamefont
  {Luttinger}}\ and\ \bibinfo {author} {\bibfnamefont {J.~C.}\ \bibnamefont
  {Ward}},\ }\href {\doibase 10.1103/PhysRev.118.1417} {\bibfield  {journal}
  {\bibinfo  {journal} {Phys. Rev.}\ }\textbf {\bibinfo {volume} {118}},\
  \bibinfo {pages} {1417} (\bibinfo {year} {1960})}\BibitemShut {NoStop}%
\bibitem [{\citenamefont {Stepanov}\ \emph {et~al.}(2019)\citenamefont
  {Stepanov}, \citenamefont {Huber}, \citenamefont {Lichtenstein},\ and\
  \citenamefont {Katsnelson}}]{Stepanov2019}%
  \BibitemOpen
  \bibfield  {author} {\bibinfo {author} {\bibfnamefont {E.~A.}\ \bibnamefont
  {Stepanov}}, \bibinfo {author} {\bibfnamefont {A.}~\bibnamefont {Huber}},
  \bibinfo {author} {\bibfnamefont {A.~I.}\ \bibnamefont {Lichtenstein}}, \
  and\ \bibinfo {author} {\bibfnamefont {M.~I.}\ \bibnamefont {Katsnelson}},\
  }\href {\doibase 10.1103/PhysRevB.99.115124} {\bibfield  {journal} {\bibinfo
  {journal} {Phys. Rev. B}\ }\textbf {\bibinfo {volume} {99}},\ \bibinfo
  {pages} {115124} (\bibinfo {year} {2019})}\BibitemShut {NoStop}%
\bibitem [{\citenamefont {Stepanov}\ \emph {et~al.}(2018)\citenamefont
  {Stepanov}, \citenamefont {Brener}, \citenamefont {Krien}, \citenamefont
  {Harland}, \citenamefont {Lichtenstein},\ and\ \citenamefont
  {Katsnelson}}]{Stepanov2018}%
  \BibitemOpen
  \bibfield  {author} {\bibinfo {author} {\bibfnamefont {E.~A.}\ \bibnamefont
  {Stepanov}}, \bibinfo {author} {\bibfnamefont {S.}~\bibnamefont {Brener}},
  \bibinfo {author} {\bibfnamefont {F.}~\bibnamefont {Krien}}, \bibinfo
  {author} {\bibfnamefont {M.}~\bibnamefont {Harland}}, \bibinfo {author}
  {\bibfnamefont {A.~I.}\ \bibnamefont {Lichtenstein}}, \ and\ \bibinfo
  {author} {\bibfnamefont {M.~I.}\ \bibnamefont {Katsnelson}},\ }\href
  {\doibase 10.1103/PhysRevLett.121.037204} {\bibfield  {journal} {\bibinfo
  {journal} {Phys. Rev. Lett.}\ }\textbf {\bibinfo {volume} {121}},\ \bibinfo
  {pages} {037204} (\bibinfo {year} {2018})}\BibitemShut {NoStop}%
\bibitem [{\citenamefont {Okamoto}\ \emph {et~al.}(2016)\citenamefont
  {Okamoto}, \citenamefont {Cavalleri},\ and\ \citenamefont
  {Mathey}}]{Okamoto2016}%
  \BibitemOpen
  \bibfield  {author} {\bibinfo {author} {\bibfnamefont {J.-i.}\ \bibnamefont
  {Okamoto}}, \bibinfo {author} {\bibfnamefont {A.}~\bibnamefont {Cavalleri}},
  \ and\ \bibinfo {author} {\bibfnamefont {L.}~\bibnamefont {Mathey}},\ }\href
  {\doibase 10.1103/PhysRevLett.117.227001} {\bibfield  {journal} {\bibinfo
  {journal} {Phys. Rev. Lett.}\ }\textbf {\bibinfo {volume} {117}},\ \bibinfo
  {pages} {227001} (\bibinfo {year} {2016})}\BibitemShut {NoStop}%
\bibitem [{\citenamefont {Okamoto}\ \emph {et~al.}(2017)\citenamefont
  {Okamoto}, \citenamefont {Hu}, \citenamefont {Cavalleri},\ and\ \citenamefont
  {Mathey}}]{Okamoto2017}%
  \BibitemOpen
  \bibfield  {author} {\bibinfo {author} {\bibfnamefont {J.-i.}\ \bibnamefont
  {Okamoto}}, \bibinfo {author} {\bibfnamefont {W.}~\bibnamefont {Hu}},
  \bibinfo {author} {\bibfnamefont {A.}~\bibnamefont {Cavalleri}}, \ and\
  \bibinfo {author} {\bibfnamefont {L.}~\bibnamefont {Mathey}},\ }\href
  {\doibase 10.1103/PhysRevB.96.144505} {\bibfield  {journal} {\bibinfo
  {journal} {Phys. Rev. B}\ }\textbf {\bibinfo {volume} {96}},\ \bibinfo
  {pages} {144505} (\bibinfo {year} {2017})}\BibitemShut {NoStop}%
\bibitem [{\citenamefont {Uemura}\ \emph {et~al.}(1988)\citenamefont {Uemura},
  \citenamefont {Emery}, \citenamefont {Moodenbaugh}, \citenamefont {Suenaga},
  \citenamefont {Johnston}, \citenamefont {Jacobson}, \citenamefont
  {Lewandowski}, \citenamefont {Brewer}, \citenamefont {Kiefl}, \citenamefont
  {Kreitzman}, \citenamefont {Luke}, \citenamefont {Riseman}, \citenamefont
  {Stronach}, \citenamefont {Kossler}, \citenamefont {Kempton}, \citenamefont
  {Yu}, \citenamefont {Opie},\ and\ \citenamefont {Schone}}]{Uemura1988}%
  \BibitemOpen
  \bibfield  {author} {\bibinfo {author} {\bibfnamefont {Y.~J.}\ \bibnamefont
  {Uemura}}, \bibinfo {author} {\bibfnamefont {V.~J.}\ \bibnamefont {Emery}},
  \bibinfo {author} {\bibfnamefont {A.~R.}\ \bibnamefont {Moodenbaugh}},
  \bibinfo {author} {\bibfnamefont {M.}~\bibnamefont {Suenaga}}, \bibinfo
  {author} {\bibfnamefont {D.~C.}\ \bibnamefont {Johnston}}, \bibinfo {author}
  {\bibfnamefont {A.~J.}\ \bibnamefont {Jacobson}}, \bibinfo {author}
  {\bibfnamefont {J.~T.}\ \bibnamefont {Lewandowski}}, \bibinfo {author}
  {\bibfnamefont {J.~H.}\ \bibnamefont {Brewer}}, \bibinfo {author}
  {\bibfnamefont {R.~F.}\ \bibnamefont {Kiefl}}, \bibinfo {author}
  {\bibfnamefont {S.~R.}\ \bibnamefont {Kreitzman}}, \bibinfo {author}
  {\bibfnamefont {G.~M.}\ \bibnamefont {Luke}}, \bibinfo {author}
  {\bibfnamefont {T.}~\bibnamefont {Riseman}}, \bibinfo {author} {\bibfnamefont
  {C.~E.}\ \bibnamefont {Stronach}}, \bibinfo {author} {\bibfnamefont {W.~J.}\
  \bibnamefont {Kossler}}, \bibinfo {author} {\bibfnamefont {J.~R.}\
  \bibnamefont {Kempton}}, \bibinfo {author} {\bibfnamefont {X.~H.}\
  \bibnamefont {Yu}}, \bibinfo {author} {\bibfnamefont {D.}~\bibnamefont
  {Opie}}, \ and\ \bibinfo {author} {\bibfnamefont {H.~E.}\ \bibnamefont
  {Schone}},\ }\href {\doibase 10.1103/PhysRevB.38.909} {\bibfield  {journal}
  {\bibinfo  {journal} {Phys. Rev. B}\ }\textbf {\bibinfo {volume} {38}},\
  \bibinfo {pages} {909} (\bibinfo {year} {1988})}\BibitemShut {NoStop}%
\bibitem [{\citenamefont {Boyce}\ \emph {et~al.}(2000)\citenamefont {Boyce},
  \citenamefont {Skinta},\ and\ \citenamefont {Lemberger}}]{Boyce2000}%
  \BibitemOpen
  \bibfield  {author} {\bibinfo {author} {\bibfnamefont {B.}~\bibnamefont
  {Boyce}}, \bibinfo {author} {\bibfnamefont {J.}~\bibnamefont {Skinta}}, \
  and\ \bibinfo {author} {\bibfnamefont {T.}~\bibnamefont {Lemberger}},\ }\href
  {\doibase https://doi.org/10.1016/S0921-4534(00)00592-X} {\bibfield
  {journal} {\bibinfo  {journal} {Physica C Supercond.}\ }\textbf {\bibinfo
  {volume} {341-348}},\ \bibinfo {pages} {561 } (\bibinfo {year}
  {2000})}\BibitemShut {NoStop}%
\bibitem [{\citenamefont {Panagopoulos}\ \emph {et~al.}(1999)\citenamefont
  {Panagopoulos}, \citenamefont {Rainford}, \citenamefont {Cooper},
  \citenamefont {Lo}, \citenamefont {Tallon}, \citenamefont {Loram},
  \citenamefont {Betouras}, \citenamefont {Wang},\ and\ \citenamefont
  {Chu}}]{Panagopoulos1999}%
  \BibitemOpen
  \bibfield  {author} {\bibinfo {author} {\bibfnamefont {C.}~\bibnamefont
  {Panagopoulos}}, \bibinfo {author} {\bibfnamefont {B.~D.}\ \bibnamefont
  {Rainford}}, \bibinfo {author} {\bibfnamefont {J.~R.}\ \bibnamefont
  {Cooper}}, \bibinfo {author} {\bibfnamefont {W.}~\bibnamefont {Lo}}, \bibinfo
  {author} {\bibfnamefont {J.~L.}\ \bibnamefont {Tallon}}, \bibinfo {author}
  {\bibfnamefont {J.~W.}\ \bibnamefont {Loram}}, \bibinfo {author}
  {\bibfnamefont {J.}~\bibnamefont {Betouras}}, \bibinfo {author}
  {\bibfnamefont {Y.~S.}\ \bibnamefont {Wang}}, \ and\ \bibinfo {author}
  {\bibfnamefont {C.~W.}\ \bibnamefont {Chu}},\ }\href {\doibase
  10.1103/PhysRevB.60.14617} {\bibfield  {journal} {\bibinfo  {journal} {Phys.
  Rev. B}\ }\textbf {\bibinfo {volume} {60}},\ \bibinfo {pages} {14617}
  (\bibinfo {year} {1999})}\BibitemShut {NoStop}%
\bibitem [{\citenamefont {Sachdev}(2011)}]{Sachdev2011}%
  \BibitemOpen
  \bibfield  {author} {\bibinfo {author} {\bibfnamefont {S.}~\bibnamefont
  {Sachdev}},\ }\href {https://books.google.de/books?id=F3IkpxwpqSgC} {\emph
  {\bibinfo {title} {Quantum Phase Transitions}}}\ (\bibinfo  {publisher}
  {Cambridge University Press},\ \bibinfo {year} {2011})\BibitemShut {NoStop}%
\bibitem [{\citenamefont {Basov}\ \emph {et~al.}(1995)\citenamefont {Basov},
  \citenamefont {Liang}, \citenamefont {Bonn}, \citenamefont {Hardy},
  \citenamefont {Dabrowski}, \citenamefont {Quijada}, \citenamefont {Tanner},
  \citenamefont {Rice}, \citenamefont {Ginsberg},\ and\ \citenamefont
  {Timusk}}]{Basov1995}%
  \BibitemOpen
  \bibfield  {author} {\bibinfo {author} {\bibfnamefont {D.~N.}\ \bibnamefont
  {Basov}}, \bibinfo {author} {\bibfnamefont {R.}~\bibnamefont {Liang}},
  \bibinfo {author} {\bibfnamefont {D.~A.}\ \bibnamefont {Bonn}}, \bibinfo
  {author} {\bibfnamefont {W.~N.}\ \bibnamefont {Hardy}}, \bibinfo {author}
  {\bibfnamefont {B.}~\bibnamefont {Dabrowski}}, \bibinfo {author}
  {\bibfnamefont {M.}~\bibnamefont {Quijada}}, \bibinfo {author} {\bibfnamefont
  {D.~B.}\ \bibnamefont {Tanner}}, \bibinfo {author} {\bibfnamefont {J.~P.}\
  \bibnamefont {Rice}}, \bibinfo {author} {\bibfnamefont {D.~M.}\ \bibnamefont
  {Ginsberg}}, \ and\ \bibinfo {author} {\bibfnamefont {T.}~\bibnamefont
  {Timusk}},\ }\href {\doibase 10.1103/PhysRevLett.74.598} {\bibfield
  {journal} {\bibinfo  {journal} {Phys. Rev. Lett.}\ }\textbf {\bibinfo
  {volume} {74}},\ \bibinfo {pages} {598} (\bibinfo {year} {1995})}\BibitemShut
  {NoStop}%
\bibitem [{\citenamefont {Homes}\ \emph {et~al.}(1995)\citenamefont {Homes},
  \citenamefont {Timusk}, \citenamefont {Bonn}, \citenamefont {Liang},\ and\
  \citenamefont {Hardy}}]{Homes1995}%
  \BibitemOpen
  \bibfield  {author} {\bibinfo {author} {\bibfnamefont {C.}~\bibnamefont
  {Homes}}, \bibinfo {author} {\bibfnamefont {T.}~\bibnamefont {Timusk}},
  \bibinfo {author} {\bibfnamefont {D.}~\bibnamefont {Bonn}}, \bibinfo {author}
  {\bibfnamefont {R.}~\bibnamefont {Liang}}, \ and\ \bibinfo {author}
  {\bibfnamefont {W.}~\bibnamefont {Hardy}},\ }\href {\doibase
  https://doi.org/10.1016/0921-4534(95)00579-X} {\bibfield  {journal} {\bibinfo
   {journal} {Physica C: Superconductivity}\ }\textbf {\bibinfo {volume}
  {254}},\ \bibinfo {pages} {265 } (\bibinfo {year} {1995})}\BibitemShut
  {NoStop}%
\bibitem [{\citenamefont {Homes}\ \emph {et~al.}(1999)\citenamefont {Homes},
  \citenamefont {Bonn}, \citenamefont {Liang}, \citenamefont {Hardy},
  \citenamefont {Basov}, \citenamefont {Timusk},\ and\ \citenamefont
  {Clayman}}]{Homes1999}%
  \BibitemOpen
  \bibfield  {author} {\bibinfo {author} {\bibfnamefont {C.~C.}\ \bibnamefont
  {Homes}}, \bibinfo {author} {\bibfnamefont {D.~A.}\ \bibnamefont {Bonn}},
  \bibinfo {author} {\bibfnamefont {R.}~\bibnamefont {Liang}}, \bibinfo
  {author} {\bibfnamefont {W.~N.}\ \bibnamefont {Hardy}}, \bibinfo {author}
  {\bibfnamefont {D.~N.}\ \bibnamefont {Basov}}, \bibinfo {author}
  {\bibfnamefont {T.}~\bibnamefont {Timusk}}, \ and\ \bibinfo {author}
  {\bibfnamefont {B.~P.}\ \bibnamefont {Clayman}},\ }\href {\doibase
  10.1103/PhysRevB.60.9782} {\bibfield  {journal} {\bibinfo  {journal} {Phys.
  Rev. B}\ }\textbf {\bibinfo {volume} {60}},\ \bibinfo {pages} {9782}
  (\bibinfo {year} {1999})}\BibitemShut {NoStop}%
\bibitem [{\citenamefont {Liu}\ \emph {et~al.}(1999)\citenamefont {Liu},
  \citenamefont {Quijada}, \citenamefont {Zibold}, \citenamefont {Yoon},
  \citenamefont {Tanner}, \citenamefont {Cao}, \citenamefont {Crow},
  \citenamefont {Berger}, \citenamefont {Margaritondo}, \citenamefont
  {Forr{\'{o}}}, \citenamefont {O}, \citenamefont {Markert}, \citenamefont
  {Kelly},\ and\ \citenamefont {Onellion}}]{Liu1999}%
  \BibitemOpen
  \bibfield  {author} {\bibinfo {author} {\bibfnamefont {H.~L.}\ \bibnamefont
  {Liu}}, \bibinfo {author} {\bibfnamefont {M.~A.}\ \bibnamefont {Quijada}},
  \bibinfo {author} {\bibfnamefont {A.~M.}\ \bibnamefont {Zibold}}, \bibinfo
  {author} {\bibfnamefont {Y.-D.}\ \bibnamefont {Yoon}}, \bibinfo {author}
  {\bibfnamefont {D.~B.}\ \bibnamefont {Tanner}}, \bibinfo {author}
  {\bibfnamefont {G.}~\bibnamefont {Cao}}, \bibinfo {author} {\bibfnamefont
  {J.~E.}\ \bibnamefont {Crow}}, \bibinfo {author} {\bibfnamefont
  {H.}~\bibnamefont {Berger}}, \bibinfo {author} {\bibfnamefont
  {G.}~\bibnamefont {Margaritondo}}, \bibinfo {author} {\bibfnamefont
  {L.}~\bibnamefont {Forr{\'{o}}}}, \bibinfo {author} {\bibfnamefont {B.-H.}\
  \bibnamefont {O}}, \bibinfo {author} {\bibfnamefont {J.~T.}\ \bibnamefont
  {Markert}}, \bibinfo {author} {\bibfnamefont {R.~J.}\ \bibnamefont {Kelly}},
  \ and\ \bibinfo {author} {\bibfnamefont {M.}~\bibnamefont {Onellion}},\
  }\href {\doibase 10.1088/0953-8984/11/1/020} {\bibfield  {journal} {\bibinfo
  {journal} {Journal of Physics: Condensed Matter}\ }\textbf {\bibinfo {volume}
  {11}},\ \bibinfo {pages} {239} (\bibinfo {year} {1999})}\BibitemShut
  {NoStop}%
\bibitem [{\citenamefont {Pereg-Barnea}\ \emph {et~al.}(2004)\citenamefont
  {Pereg-Barnea}, \citenamefont {Turner}, \citenamefont {Harris}, \citenamefont
  {Mullins}, \citenamefont {Bobowski}, \citenamefont {Raudsepp}, \citenamefont
  {Liang}, \citenamefont {Bonn},\ and\ \citenamefont
  {Hardy}}]{Pereg-Barnea2004}%
  \BibitemOpen
  \bibfield  {author} {\bibinfo {author} {\bibfnamefont {T.}~\bibnamefont
  {Pereg-Barnea}}, \bibinfo {author} {\bibfnamefont {P.~J.}\ \bibnamefont
  {Turner}}, \bibinfo {author} {\bibfnamefont {R.}~\bibnamefont {Harris}},
  \bibinfo {author} {\bibfnamefont {G.~K.}\ \bibnamefont {Mullins}}, \bibinfo
  {author} {\bibfnamefont {J.~S.}\ \bibnamefont {Bobowski}}, \bibinfo {author}
  {\bibfnamefont {M.}~\bibnamefont {Raudsepp}}, \bibinfo {author}
  {\bibfnamefont {R.}~\bibnamefont {Liang}}, \bibinfo {author} {\bibfnamefont
  {D.~A.}\ \bibnamefont {Bonn}}, \ and\ \bibinfo {author} {\bibfnamefont
  {W.~N.}\ \bibnamefont {Hardy}},\ }\href {\doibase 10.1103/PhysRevB.69.184513}
  {\bibfield  {journal} {\bibinfo  {journal} {Phys. Rev. B}\ }\textbf {\bibinfo
  {volume} {69}},\ \bibinfo {pages} {184513} (\bibinfo {year}
  {2004})}\BibitemShut {NoStop}%
\bibitem [{\citenamefont {Homes}\ \emph {et~al.}(2004)\citenamefont {Homes},
  \citenamefont {Dordevic}, \citenamefont {Strongin}, \citenamefont {Bonn},
  \citenamefont {Liang}, \citenamefont {Hardy}, \citenamefont {Komiya},
  \citenamefont {Ando}, \citenamefont {Yu}, \citenamefont {Kaneko},
  \citenamefont {Zhao}, \citenamefont {Greven}, \citenamefont {Basov},\ and\
  \citenamefont {Timusk}}]{Homes2004}%
  \BibitemOpen
  \bibfield  {author} {\bibinfo {author} {\bibfnamefont {C.~C.}\ \bibnamefont
  {Homes}}, \bibinfo {author} {\bibfnamefont {S.~V.}\ \bibnamefont {Dordevic}},
  \bibinfo {author} {\bibfnamefont {M.}~\bibnamefont {Strongin}}, \bibinfo
  {author} {\bibfnamefont {D.~A.}\ \bibnamefont {Bonn}}, \bibinfo {author}
  {\bibfnamefont {R.}~\bibnamefont {Liang}}, \bibinfo {author} {\bibfnamefont
  {W.~N.}\ \bibnamefont {Hardy}}, \bibinfo {author} {\bibfnamefont
  {S.}~\bibnamefont {Komiya}}, \bibinfo {author} {\bibfnamefont
  {Y.}~\bibnamefont {Ando}}, \bibinfo {author} {\bibfnamefont {G.}~\bibnamefont
  {Yu}}, \bibinfo {author} {\bibfnamefont {N.}~\bibnamefont {Kaneko}}, \bibinfo
  {author} {\bibfnamefont {X.}~\bibnamefont {Zhao}}, \bibinfo {author}
  {\bibfnamefont {M.}~\bibnamefont {Greven}}, \bibinfo {author} {\bibfnamefont
  {D.~N.}\ \bibnamefont {Basov}}, \ and\ \bibinfo {author} {\bibfnamefont
  {T.}~\bibnamefont {Timusk}},\ }\href {https://doi.org/10.1038/nature02673}
  {\bibfield  {journal} {\bibinfo  {journal} {Nature}\ }\textbf {\bibinfo
  {volume} {430}},\ \bibinfo {pages} {539 EP } (\bibinfo {year}
  {2004})}\BibitemShut {NoStop}%
\bibitem [{\citenamefont {Dordevic}\ \emph {et~al.}(2013)\citenamefont
  {Dordevic}, \citenamefont {Basov},\ and\ \citenamefont
  {Homes}}]{Dordevic2013}%
  \BibitemOpen
  \bibfield  {author} {\bibinfo {author} {\bibfnamefont {S.~V.}\ \bibnamefont
  {Dordevic}}, \bibinfo {author} {\bibfnamefont {D.~N.}\ \bibnamefont {Basov}},
  \ and\ \bibinfo {author} {\bibfnamefont {C.~C.}\ \bibnamefont {Homes}},\
  }\href {https://doi.org/10.1038/srep01713} {\bibfield  {journal} {\bibinfo
  {journal} {Scientific Reports}\ }\textbf {\bibinfo {volume} {3}},\ \bibinfo
  {pages} {1713 EP } (\bibinfo {year} {2013})},\ \bibinfo {note}
  {article}\BibitemShut {NoStop}%
\bibitem [{\citenamefont {Kamal}\ \emph {et~al.}(1998)\citenamefont {Kamal},
  \citenamefont {Liang}, \citenamefont {Hosseini}, \citenamefont {Bonn},\ and\
  \citenamefont {Hardy}}]{Kamal1998}%
  \BibitemOpen
  \bibfield  {author} {\bibinfo {author} {\bibfnamefont {S.}~\bibnamefont
  {Kamal}}, \bibinfo {author} {\bibfnamefont {R.}~\bibnamefont {Liang}},
  \bibinfo {author} {\bibfnamefont {A.}~\bibnamefont {Hosseini}}, \bibinfo
  {author} {\bibfnamefont {D.~A.}\ \bibnamefont {Bonn}}, \ and\ \bibinfo
  {author} {\bibfnamefont {W.~N.}\ \bibnamefont {Hardy}},\ }\href {\doibase
  10.1103/PhysRevB.58.R8933} {\bibfield  {journal} {\bibinfo  {journal} {Phys.
  Rev. B}\ }\textbf {\bibinfo {volume} {58}},\ \bibinfo {pages} {R8933}
  (\bibinfo {year} {1998})}\BibitemShut {NoStop}%
\bibitem [{\citenamefont {Nelson}\ and\ \citenamefont
  {Kosterlitz}(1977)}]{Nelson1977}%
  \BibitemOpen
  \bibfield  {author} {\bibinfo {author} {\bibfnamefont {D.~R.}\ \bibnamefont
  {Nelson}}\ and\ \bibinfo {author} {\bibfnamefont {J.~M.}\ \bibnamefont
  {Kosterlitz}},\ }\href {\doibase 10.1103/PhysRevLett.39.1201} {\bibfield
  {journal} {\bibinfo  {journal} {Phys. Rev. Lett.}\ }\textbf {\bibinfo
  {volume} {39}},\ \bibinfo {pages} {1201} (\bibinfo {year}
  {1977})}\BibitemShut {NoStop}%
\bibitem [{\citenamefont {Mermin}\ and\ \citenamefont
  {Wagner}(1966)}]{Mermin1966}%
  \BibitemOpen
  \bibfield  {author} {\bibinfo {author} {\bibfnamefont {N.~D.}\ \bibnamefont
  {Mermin}}\ and\ \bibinfo {author} {\bibfnamefont {H.}~\bibnamefont
  {Wagner}},\ }\href {\doibase 10.1103/PhysRevLett.17.1133} {\bibfield
  {journal} {\bibinfo  {journal} {Phys. Rev. Lett.}\ }\textbf {\bibinfo
  {volume} {17}},\ \bibinfo {pages} {1133} (\bibinfo {year}
  {1966})}\BibitemShut {NoStop}%
\bibitem [{\citenamefont {Benfatto}\ \emph {et~al.}(2007)\citenamefont
  {Benfatto}, \citenamefont {Castellani},\ and\ \citenamefont
  {Giamarchi}}]{Benfatto2007}%
  \BibitemOpen
  \bibfield  {author} {\bibinfo {author} {\bibfnamefont {L.}~\bibnamefont
  {Benfatto}}, \bibinfo {author} {\bibfnamefont {C.}~\bibnamefont
  {Castellani}}, \ and\ \bibinfo {author} {\bibfnamefont {T.}~\bibnamefont
  {Giamarchi}},\ }\href {\doibase 10.1103/PhysRevLett.98.117008} {\bibfield
  {journal} {\bibinfo  {journal} {Phys. Rev. Lett.}\ }\textbf {\bibinfo
  {volume} {98}},\ \bibinfo {pages} {117008} (\bibinfo {year}
  {2007})}\BibitemShut {NoStop}%
\bibitem [{\citenamefont {Sordi}\ \emph {et~al.}(2010)\citenamefont {Sordi},
  \citenamefont {Haule},\ and\ \citenamefont {Tremblay}}]{Sordi2010}%
  \BibitemOpen
  \bibfield  {author} {\bibinfo {author} {\bibfnamefont {G.}~\bibnamefont
  {Sordi}}, \bibinfo {author} {\bibfnamefont {K.}~\bibnamefont {Haule}}, \ and\
  \bibinfo {author} {\bibfnamefont {A.-M.~S.}\ \bibnamefont {Tremblay}},\
  }\href {\doibase 10.1103/PhysRevLett.104.226402} {\bibfield  {journal}
  {\bibinfo  {journal} {Phys. Rev. Lett.}\ }\textbf {\bibinfo {volume} {104}},\
  \bibinfo {pages} {226402} (\bibinfo {year} {2010})}\BibitemShut {NoStop}%
\bibitem [{\citenamefont {Kancharla}\ \emph {et~al.}(2008)\citenamefont
  {Kancharla}, \citenamefont {Kyung}, \citenamefont {S\'en\'echal},
  \citenamefont {Civelli}, \citenamefont {Capone}, \citenamefont {Kotliar},\
  and\ \citenamefont {Tremblay}}]{Kancharla2008}%
  \BibitemOpen
  \bibfield  {author} {\bibinfo {author} {\bibfnamefont {S.~S.}\ \bibnamefont
  {Kancharla}}, \bibinfo {author} {\bibfnamefont {B.}~\bibnamefont {Kyung}},
  \bibinfo {author} {\bibfnamefont {D.}~\bibnamefont {S\'en\'echal}}, \bibinfo
  {author} {\bibfnamefont {M.}~\bibnamefont {Civelli}}, \bibinfo {author}
  {\bibfnamefont {M.}~\bibnamefont {Capone}}, \bibinfo {author} {\bibfnamefont
  {G.}~\bibnamefont {Kotliar}}, \ and\ \bibinfo {author} {\bibfnamefont
  {A.-M.~S.}\ \bibnamefont {Tremblay}},\ }\href {\doibase
  10.1103/PhysRevB.77.184516} {\bibfield  {journal} {\bibinfo  {journal} {Phys.
  Rev. B}\ }\textbf {\bibinfo {volume} {77}},\ \bibinfo {pages} {184516}
  (\bibinfo {year} {2008})}\BibitemShut {NoStop}%
\bibitem [{\citenamefont {Chen}\ \emph {et~al.}(2005)\citenamefont {Chen},
  \citenamefont {Stajic}, \citenamefont {Tan},\ and\ \citenamefont
  {Levin}}]{Chen2005}%
  \BibitemOpen
  \bibfield  {author} {\bibinfo {author} {\bibfnamefont {Q.}~\bibnamefont
  {Chen}}, \bibinfo {author} {\bibfnamefont {J.}~\bibnamefont {Stajic}},
  \bibinfo {author} {\bibfnamefont {S.}~\bibnamefont {Tan}}, \ and\ \bibinfo
  {author} {\bibfnamefont {K.}~\bibnamefont {Levin}},\ }\href {\doibase
  https://doi.org/10.1016/j.physrep.2005.02.005} {\bibfield  {journal}
  {\bibinfo  {journal} {Physics Reports}\ }\textbf {\bibinfo {volume} {412}},\
  \bibinfo {pages} {1 } (\bibinfo {year} {2005})}\BibitemShut {NoStop}%
\bibitem [{\citenamefont {Damascelli}\ \emph {et~al.}(2003)\citenamefont
  {Damascelli}, \citenamefont {Hussain},\ and\ \citenamefont
  {Shen}}]{Damascelli2003}%
  \BibitemOpen
  \bibfield  {author} {\bibinfo {author} {\bibfnamefont {A.}~\bibnamefont
  {Damascelli}}, \bibinfo {author} {\bibfnamefont {Z.}~\bibnamefont {Hussain}},
  \ and\ \bibinfo {author} {\bibfnamefont {Z.-X.}\ \bibnamefont {Shen}},\
  }\href {\doibase 10.1103/RevModPhys.75.473} {\bibfield  {journal} {\bibinfo
  {journal} {Rev. Mod. Phys.}\ }\textbf {\bibinfo {volume} {75}},\ \bibinfo
  {pages} {473} (\bibinfo {year} {2003})}\BibitemShut {NoStop}%
\bibitem [{\citenamefont {Zeh}\ \emph {et~al.}(1990)\citenamefont {Zeh},
  \citenamefont {Ri}, \citenamefont {Kober}, \citenamefont {Huebener},
  \citenamefont {Ustinov}, \citenamefont {Mannhart}, \citenamefont {Gross},\
  and\ \citenamefont {Gupta}}]{Zeh1990}%
  \BibitemOpen
  \bibfield  {author} {\bibinfo {author} {\bibfnamefont {M.}~\bibnamefont
  {Zeh}}, \bibinfo {author} {\bibfnamefont {H.-C.}\ \bibnamefont {Ri}},
  \bibinfo {author} {\bibfnamefont {F.}~\bibnamefont {Kober}}, \bibinfo
  {author} {\bibfnamefont {R.~P.}\ \bibnamefont {Huebener}}, \bibinfo {author}
  {\bibfnamefont {A.~V.}\ \bibnamefont {Ustinov}}, \bibinfo {author}
  {\bibfnamefont {J.}~\bibnamefont {Mannhart}}, \bibinfo {author}
  {\bibfnamefont {R.}~\bibnamefont {Gross}}, \ and\ \bibinfo {author}
  {\bibfnamefont {A.}~\bibnamefont {Gupta}},\ }\href {\doibase
  10.1103/PhysRevLett.64.3195} {\bibfield  {journal} {\bibinfo  {journal}
  {Phys. Rev. Lett.}\ }\textbf {\bibinfo {volume} {64}},\ \bibinfo {pages}
  {3195} (\bibinfo {year} {1990})}\BibitemShut {NoStop}%
\bibitem [{\citenamefont {Ri}\ \emph {et~al.}(1994)\citenamefont {Ri},
  \citenamefont {Gross}, \citenamefont {Gollnik}, \citenamefont {Beck},
  \citenamefont {Huebener}, \citenamefont {Wagner},\ and\ \citenamefont
  {Adrian}}]{Ri1994}%
  \BibitemOpen
  \bibfield  {author} {\bibinfo {author} {\bibfnamefont {H.-C.}\ \bibnamefont
  {Ri}}, \bibinfo {author} {\bibfnamefont {R.}~\bibnamefont {Gross}}, \bibinfo
  {author} {\bibfnamefont {F.}~\bibnamefont {Gollnik}}, \bibinfo {author}
  {\bibfnamefont {A.}~\bibnamefont {Beck}}, \bibinfo {author} {\bibfnamefont
  {R.~P.}\ \bibnamefont {Huebener}}, \bibinfo {author} {\bibfnamefont
  {P.}~\bibnamefont {Wagner}}, \ and\ \bibinfo {author} {\bibfnamefont
  {H.}~\bibnamefont {Adrian}},\ }\href {\doibase 10.1103/PhysRevB.50.3312}
  {\bibfield  {journal} {\bibinfo  {journal} {Phys. Rev. B}\ }\textbf {\bibinfo
  {volume} {50}},\ \bibinfo {pages} {3312} (\bibinfo {year}
  {1994})}\BibitemShut {NoStop}%
\bibitem [{\citenamefont {Wang}\ \emph {et~al.}(2006)\citenamefont {Wang},
  \citenamefont {Li},\ and\ \citenamefont {Ong}}]{Wang2006}%
  \BibitemOpen
  \bibfield  {author} {\bibinfo {author} {\bibfnamefont {Y.}~\bibnamefont
  {Wang}}, \bibinfo {author} {\bibfnamefont {L.}~\bibnamefont {Li}}, \ and\
  \bibinfo {author} {\bibfnamefont {N.~P.}\ \bibnamefont {Ong}},\ }\href
  {\doibase 10.1103/PhysRevB.73.024510} {\bibfield  {journal} {\bibinfo
  {journal} {Phys. Rev. B}\ }\textbf {\bibinfo {volume} {73}},\ \bibinfo
  {pages} {024510} (\bibinfo {year} {2006})}\BibitemShut {NoStop}%
\bibitem [{\citenamefont {Kivelson}\ \emph {et~al.}(2003)\citenamefont
  {Kivelson}, \citenamefont {Bindloss}, \citenamefont {Fradkin}, \citenamefont
  {Oganesyan}, \citenamefont {Tranquada}, \citenamefont {Kapitulnik},\ and\
  \citenamefont {Howald}}]{Kivelson2003}%
  \BibitemOpen
  \bibfield  {author} {\bibinfo {author} {\bibfnamefont {S.~A.}\ \bibnamefont
  {Kivelson}}, \bibinfo {author} {\bibfnamefont {I.~P.}\ \bibnamefont
  {Bindloss}}, \bibinfo {author} {\bibfnamefont {E.}~\bibnamefont {Fradkin}},
  \bibinfo {author} {\bibfnamefont {V.}~\bibnamefont {Oganesyan}}, \bibinfo
  {author} {\bibfnamefont {J.~M.}\ \bibnamefont {Tranquada}}, \bibinfo {author}
  {\bibfnamefont {A.}~\bibnamefont {Kapitulnik}}, \ and\ \bibinfo {author}
  {\bibfnamefont {C.}~\bibnamefont {Howald}},\ }\href {\doibase
  10.1103/RevModPhys.75.1201} {\bibfield  {journal} {\bibinfo  {journal} {Rev.
  Mod. Phys.}\ }\textbf {\bibinfo {volume} {75}},\ \bibinfo {pages} {1201}
  (\bibinfo {year} {2003})}\BibitemShut {NoStop}%
\bibitem [{\citenamefont {Wollny}\ and\ \citenamefont
  {Vojta}(2009)}]{Wollny2009}%
  \BibitemOpen
  \bibfield  {author} {\bibinfo {author} {\bibfnamefont {A.}~\bibnamefont
  {Wollny}}\ and\ \bibinfo {author} {\bibfnamefont {M.}~\bibnamefont {Vojta}},\
  }\href {\doibase 10.1103/PhysRevB.80.132504} {\bibfield  {journal} {\bibinfo
  {journal} {Phys. Rev. B}\ }\textbf {\bibinfo {volume} {80}},\ \bibinfo
  {pages} {132504} (\bibinfo {year} {2009})}\BibitemShut {NoStop}%
\bibitem [{\citenamefont {Rajasekaran}\ \emph {et~al.}(2018)\citenamefont
  {Rajasekaran}, \citenamefont {Okamoto}, \citenamefont {Mathey}, \citenamefont
  {Fechner}, \citenamefont {Thampy}, \citenamefont {Gu},\ and\ \citenamefont
  {Cavalleri}}]{Rajasekaran2018}%
  \BibitemOpen
  \bibfield  {author} {\bibinfo {author} {\bibfnamefont {S.}~\bibnamefont
  {Rajasekaran}}, \bibinfo {author} {\bibfnamefont {J.}~\bibnamefont
  {Okamoto}}, \bibinfo {author} {\bibfnamefont {L.}~\bibnamefont {Mathey}},
  \bibinfo {author} {\bibfnamefont {M.}~\bibnamefont {Fechner}}, \bibinfo
  {author} {\bibfnamefont {V.}~\bibnamefont {Thampy}}, \bibinfo {author}
  {\bibfnamefont {G.~D.}\ \bibnamefont {Gu}}, \ and\ \bibinfo {author}
  {\bibfnamefont {A.}~\bibnamefont {Cavalleri}},\ }\href {\doibase
  10.1126/science.aan3438} {\bibfield  {journal} {\bibinfo  {journal}
  {Science}\ }\textbf {\bibinfo {volume} {359}},\ \bibinfo {pages} {575}
  (\bibinfo {year} {2018})}\BibitemShut {NoStop}%
\bibitem [{\citenamefont {Georges}\ \emph {et~al.}(1996)\citenamefont
  {Georges}, \citenamefont {Kotliar}, \citenamefont {Krauth},\ and\
  \citenamefont {Rozenberg}}]{Georges1996}%
  \BibitemOpen
  \bibfield  {author} {\bibinfo {author} {\bibfnamefont {A.}~\bibnamefont
  {Georges}}, \bibinfo {author} {\bibfnamefont {G.}~\bibnamefont {Kotliar}},
  \bibinfo {author} {\bibfnamefont {W.}~\bibnamefont {Krauth}}, \ and\ \bibinfo
  {author} {\bibfnamefont {M.~J.}\ \bibnamefont {Rozenberg}},\ }\href {\doibase
  10.1103/RevModPhys.68.13} {\bibfield  {journal} {\bibinfo  {journal} {Rev.
  Mod. Phys.}\ }\textbf {\bibinfo {volume} {68}},\ \bibinfo {pages} {13}
  (\bibinfo {year} {1996})}\BibitemShut {NoStop}%
\end{thebibliography}%

\appendix

\section{Tightbinding model\label{app:tightbinding}}
In most strong-coupling calculations on copper-oxides theoreticians use the single band Hubbard model as the main features are believed to exist in the square lattice symmetry. However, starting density functional calculations one can also integrate out the bands at energies distant from Fermi level and obtain an effective one-band model, that has been done for YBCO\cite{Andersen1995}. At this point we note that the complicated structure of YBCO which consists of bilayers with the intra-bilayer hopping of the order of $0.65$ in units of $t$ results in a splitting between bonding and anti-bonding bands with the value of the splitting being much larger than the individual bandwidth of each of those. This is the reason why it is possible in the first approximation to consider an effective one (anti-bonding) band model. In this section we compare the effects of the bandstructures on the dSC stiffness also for a simple perpendicular hopping.

\begin{figure}
  \centering
  \includegraphics{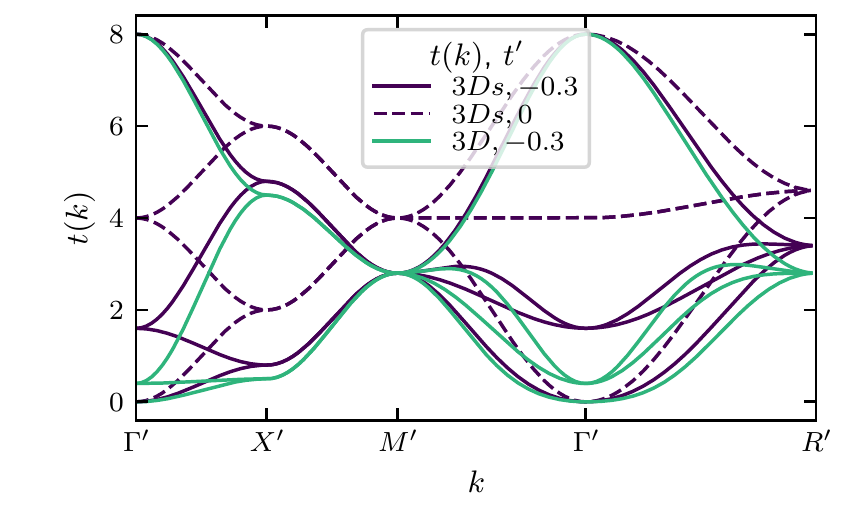}
  \caption{Electronic tight binding bandstructures of the different hopping lattices $3D$ and $3D^\ast$ and next-nearest neighbor hoppings $t^\prime$. $k$ is the reciprocal lattice vector in the reduced Brillouin zone. The four bands correspond to four sites within the two-by-two plaquette unit cell. With our choice of cluster the reduced Brillouin zone has the same shape as the original Brillouin zone of the square lattice. Thus we label the high symmetry points accordingly, but with a prime. $\Gamma^\prime = (0,0,0)$, $X^\prime = (1,0,0)$, $M^\prime = (1,1,0)$, $R^\prime = (1,1,1)$ in units of half of the reduced Brillouin zone.}
  \label{fig:tb}
\end{figure}
The 2D dispersion is that of the square lattice
\begin{equation}
  \label{eq:apptbtk2d}
  t^{2D}(k)= 2t \left( \cos(k_x) + \cos(k_y) \right) - 4t^\prime \cos(k_x) \cos(k_y),
\end{equation}
then, for three dimensions we can compare a simple perpendicular hopping model ($3Ds$)
\begin{equation}
  \label{eq:apptbtk3ds}
  t^{3Ds}(k) = t^{2D}(k) + 2 t_\perp \cos(k_z)
\end{equation}
with a more elaborated projection\cite{Andersen1995} ($3D$)
\begin{equation}
  \label{eq:apptbtk3d}
  t^{3D}(k) = t^{2D}(k) + 2 \frac{t_\perp}{4} \left( \cos(k_x) - \cos(k_y) \right)^2 \cos(k_z).
\end{equation}
In \refeq{eq:apptbtk2d} to \refeq{eq:apptbtk3d} $k$ is in the full Brillouin zone. For a cluster formulation $k$ has to be in the reduced Brillouin zone according to the reduced translational symmetry. The bandstructure shown in \reffig{fig:tb} has four bands corresponding to the cluster of four sites. The hopping matrices $t_r$ of plaquette translations $r$ for the $3D$ model read
\begin{widetext}
\begin{gather}
  t_{(0,0,0)} = 
  \begin{pmatrix} 0 & t & t & t^\prime \\ t & 0 & t^\prime & t \\ t & t^\prime & 0 & t \\ t^\prime & t & t & 0\end{pmatrix},\,
  t_{(1,0,0)} =
  \begin{pmatrix} 0 & t & 0 & t^\prime \\ 0 & 0 & 0 & 0 \\ 0 & t^\prime & 0 & t \\ 0 & 0 & 0 & 0\end{pmatrix},\,
  t_{(1,1,0)} =
  \begin{pmatrix} 0 & 0 & 0 & t^\prime \\ 0 & 0 & 0 & 0 \\ 0 & 0 & 0 & 0 \\ 0 & 0 & 0 & 0\end{pmatrix},\,
  t_{(0,1,0)} =
  \begin{pmatrix} 0 & 0 & t & t^\prime \\ 0 & 0 & t^\prime & t \\ 0 & 0 & 0 & 0 \\ 0 & 0 & 0 & 0\end{pmatrix},\nonumber\\
  t_{(-1,1,0)} =
  \begin{pmatrix} 0 & 0 & 0 & 0 \\ 0 & 0 & t^\prime & 0 \\ 0 & 0 & 0 & 0 \\ 0 & 0 & 0 & 0\end{pmatrix},\,
  t_{(0,0,1)} =
  \begin{pmatrix} t_0 & 0 & 0 & t_2 \\ 0 & t_0 & t_2 & 0 \\ 0 & t_2 & t_0 & 0 \\ t_2 & 0 & 0 & t_0\end{pmatrix},\,
  t_{(1,0,1)} = t_{(1,0,-1)} =
  \begin{pmatrix} t_1 & 0 & 0 & t_2 \\ 0 & t_1 & 0 & 0 \\ 0 & t_2 & t_1 & 0 \\ 0 & 0 & 0 & t_1\end{pmatrix},\\
  t_{(0,1,1)} = t_{(0,1,-1)} =
  \begin{pmatrix} t_1 & 0 & 0 & t_2 \\ 0 & t_1 & t_2 & 0 \\ 0 & 0 & t_1 & 0 \\ 0 & 0 & 0 & t_1\end{pmatrix},\,
  t_{(1,1,1)} = t_{(1,1,-1)} =
  \begin{pmatrix} 0 & 0 & 0 & t_2 \\ 0 & 0 & 0 & 0 \\ 0 & 0 & 0 & 0 \\ 0 & 0 & 0 & 0\end{pmatrix},\,
  t_{(-1,1,1)} = t_{(-1,1,-1)} =
  \begin{pmatrix} 0 & 0 & 0 & 0 \\ 0 & 0 & t_2 & 0 \\ 0 & 0 & 0 & 0 \\ 0 & 0 & 0 & 0\end{pmatrix},\nonumber
\end{gather}
\end{widetext}
with $t_0 = t_\perp/4$, $t_1 = t_\perp/16$, $t_2 = t_\perp/8$ and $t_{-r} = t^\intercal_{r}$. The entries correspond to the clustersites, labeled according to \reffig{fig:illu}. 

\section{Green functions in CDMFT\label{app:cdmft}}
We solve the CDMFT\cite{Georges1996,Lichtenstein2000,Kotliar2001} equation
\begin{gather}
  \label{eq:cdmft}
    \mathcal{G}^{-1}(i\omega_n) = \left(\sum_{k} G(i\omega_n, k)\right)^{-1} + \Sigma(i\omega_n),\\
    G^{-1}(i\omega_n, k) = i\omega_n + \mu - t(k) + \Sigma(i\omega_n)\label{eq:cdmft2}
\end{gather}
with the lattice dispersion of the reduced Brillouin zone $t(k)$ numerically\cite{Parcollet2015,Seth2016} and obtain the self-consistent local lattice Green function that is the first term on the r.h.s. of \refeq{eq:cdmft}. The chemical potential for a certain doping can be found by solving only \refeq{eq:cdmft2} iteratively. But this is an additional quantity that has to converge with the CDMFT cycles. To make the CDMFT more efficient in that regard, we set a certain chemical potential $\mu$ as the parameter rather than the doping. This gives a non-uniform mesh in the temperature-doping phase diagram and requires a postprocessing of two-dimensional interpolation. CDMFT maps the lattice problem to a multiorbital Anderson impurity model, in that the different orbitals also represent the sites of the cluster. The Anderson impurity model of arbitrary local interactions can be solved exactly by the use of the continous-time quantum Monte-Carlo method (CTHYB). The bath of that model is dynamical and so is the mean-field of CDMFT. But the temporal correlations exist only locally, i.e. on the cluster. Therefore the self-energy between clusters vanishes.

Using the symmetry of the plaquette, the local Green function has the blockstructure
\begin{equation}
  \label{eq:gblockstruc}
  G_{loc} =
  \begin{pmatrix}
    G_\Gamma & & & \\
    & G_X & &  \\
    & & G_Y & \\
    & & & G_M
  \end{pmatrix}
\end{equation}
where we labeled the plaquette orbitals according to the same transformation properties of the high-symmetry points of the Brillouin zone of the squarelattice. The transformation from site-space to plaquette orbitals is a unitary transformation with
\begin{equation}
  \label{eq:sitetransf}
  U = \frac{1}{2}
  \begin{pmatrix}
    1&1&1&1\\
    1&-1&1&-1\\
    1&1&-1&-1\\
    1&-1&-1&1
  \end{pmatrix}
\end{equation}
In principle antiferromagnetic order can also be considered, but it would reduce the blockstructure of \refeq{eq:gblockstruc} and will be computationally more expensive.

In our CDMFT approximation the self-energy exists only within the cluster and not between clusters. In order to obtain the lattice Green function one could try to interpolate the many-body correlations between the clusters. This procedure is ambiguous. Following the idea of strong correlations within the plaquette being crucial, we do not interpolate the self-energy. The locality of the self-energy is required for the applicability of the local force theorem. In that aspect the CDMFT we use and the local force theorem are perfectly compatible as they make the same assumptions. Therefore the lattice Green function reads
\begin{equation}
  \label{eq:glat}
  G(i\omega_n, r) = \frac{1}{N_{k}} \sum_{k} \frac{e^{ikr}}{i\omega_n +\mu - t(k) - \Sigma(i\omega_n)},
\end{equation}
where $r$ are cluster-translations and $i\omega$, $\mu$, $t(k)$ and $\Sigma(i\omega_n)$ are matrices in Nambu plaquette-orbital or site-basis. $k$ is in the reduced Brillouin zone according to plaquette translations. For the CDMFT calculations we use $1025$ Matsubara frequencies, $64$ $k$-points per dimension, $192\times 10^5$ Monte-Carlo (MC) measurements, $200$ updates per MC measurement and $3\times 10^3$ MC warm-up cycles. The number of Legendre-coefficients for the representation of the Green function, that we measure in the Monte-Carlo process, depends mostly on the temperature. A reasonable range for our calculations is $50$-$150$. During the CDMFT loops we perform partial updates of the self-energy using a mixing parameter of $0.5$. For the dSC symmetry breaking we initialized the CDMFT cycles with a symmetry breaking seed in the self-energy.

\begin{figure}
  \centering
  \includegraphics{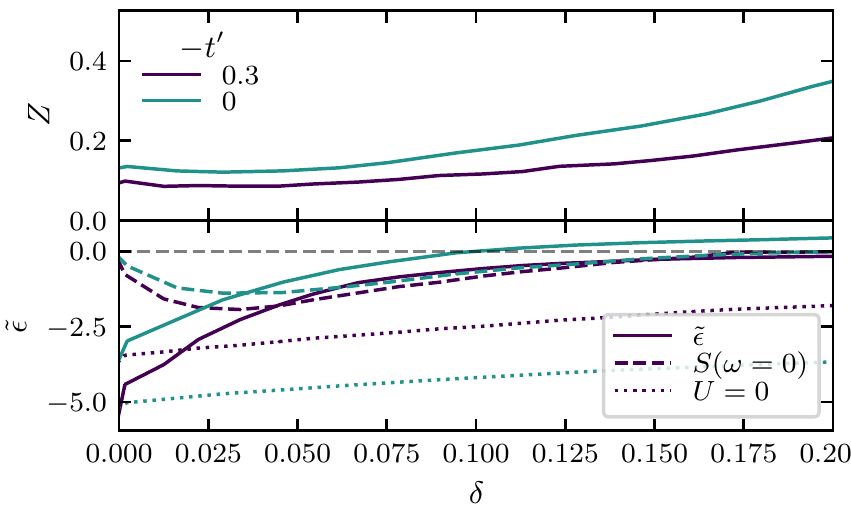}
  \caption{The quasiparticle residue $Z$ (top) and energy $\tilde{\epsilon}$ of $k=X$ (bottom) as functions of the hole-doping $\delta$. The non-interaction quasiparticle energy ($U=0$) and the anomalous part $S$ are also shown (bottom). ($T=1/52$, $t_\perp = 0$).}
  \label{fig:z_eps}
\end{figure}
A success of the DMFT is the description of the Mott insulator, an insulator of odd-integer filling, that is gapped by local correlation effects induced by $U$. It can be characterized by the vanishing quasiparticle residue
\begin{equation}
  \label{eq:z}
  Z_k^{-1} = 1-\frac{\partial \Re \Sigma_k(\omega)}{\partial \omega}\Big|_{\omega = 0}
\end{equation}
of that $k$-point, whose energy corresponds to the Fermi energy and at $T=0$. Furthermore we have the quasiparticle energy
\begin{equation}
  \label{eq:eps}
  \tilde{\epsilon}_X= -\mu -4t^\prime + \Re \Sigma_X(\omega = 0)
\end{equation}
whose zeros can indicate the Lifshitz transition\cite{Wu2018,Braganca2018}, at that the Fermi surface turns from particle-like to hole-like. We present these quantitites for symmetry broken solutions. Thus there is a gap and no quasiparticles. However, assuming that the feedback of a finite anomalous self-energy $S$ on the normal parts is small and extract information on the underlying electron quasiparticles and correlations.

The quasiparticle get significantly renormalized close to half-filling resembling Mottness, see \reffig{fig:z_eps}. The Mott insulator is known to be connected to metallic states by a first order transition.\cite{Georges1996} The anomalous part of the self-energy $S$ makes an essentail contribution to the Josephson coupling and the dSC stiffness. It can be seen in \reffig{fig:z_eps} that it becomes small at small frequencies around $\delta \sim 0.15$ at $T\sim 0.2$.

\section{Gauge invariance and its consequences\label{app:gi}}
Sum-rules express correlations of certain transitions in terms of sums over other transitions. We derive a set of sum-rules starting from the Dyson equation. In this section we work in the Nambu-space (omitting the spin labes for convenience), but the quantities can still be matrices of other subspaces. Therefore we have
\begin{align}
  \label{eq:sr1}
  \begin{split}
    G &= \begin{pmatrix} G^p && \fan \\ \fan && G^h \end{pmatrix}, \\
    G^{-1}_0 &= \begin{pmatrix} \left(G^p_0\right)^{-1} && 0 \\ 0 && \left(G^h_0\right)^{-1} \end{pmatrix}, \\
    \Sigma &= \begin{pmatrix} \Sigma^p && S \\ S && \Sigma^h \end{pmatrix}.
  \end{split}
\end{align}
We temporarily switch to the bonding-/antibonding ($+,-$) basis
\begin{equation}
  \label{eq:sr2}
  2G^+ = G^p + G^h,\quad 2G^- = G^p - G^h
\end{equation}
and for $\Sigma$ and $G_0$ accordingly. We expand the correlation functions in Pauli matices:
\begin{align}
  \label{eq:sr3}
  \begin{split}
    G &= G^+ \mathbbm{1} + \left(\fan,\,0,\,G^-\right) \bm{\sigma},\\
    \Sigma &= \Sigma^+ \mathbbm{1} + \left(S,\,0,\,\Sigma^-\right) \bm{\sigma},\\
    G_0 &= G_0^+ \mathbbm{1} + \left(0,\,0,\,G_0^-\right) \bm{\sigma}.
  \end{split}
\end{align}
The Dyson equation then reads
\begin{equation}
  \label{eq:sr4}
  G^{-1} = \left(G^+_0 - \Sigma^+\right)\mathbbm{1} + \left(S,\, 0,\, G^-_0 - \Sigma^-\right)\bm{\sigma}.
\end{equation}
The identity
\begin{equation}
  \label{eq:sr5}
  G G^{-1} = \mathbbm{1}
\end{equation}
leads to a set of four equations:
\begin{align}
  \mathbbm{1} &= G^+\left(G_0^+-\Sigma^+\right) - \fan S + G^- \left(G^-_0 - \Sigma^-\right), \label{eq:pauli0} \\
  0 &= \fan \left(G_0^+-\Sigma^+\right) - G^+ S, \label{eq:pauli1} \\
  0 &= \fan \left(G_0^--\Sigma^-\right) + G^- S, \label{eq:pauli2} \\
  0 &= G^+ \left(G_0^--\Sigma^-\right) + G^- \left(G_0^+-\Sigma^+\right). \label{eq:pauli3}
\end{align}
From \refeq{eq:pauli1} and \refeq{eq:pauli2} directly follows
\begin{align}
  \left(G_0^+-\Sigma^+\right) &= \fan^{-1}G^+ S,\\
  \left(G_0^--\Sigma^-\right) &= -\fan^{-1}G^- S,
\end{align}
which we insert in \refeq{eq:pauli0} also backtransforming the $(+,-)$ basis,
\begin{equation}
  \label{eq:sr6}
  \mathbbm{1} = - \fan S + \frac{1}{2} \left( G^p \fan^{-1} G^h S + G^h \fan^{-1} G^p S\right).
\end{equation}
Furthermore we insert \refeq{eq:pauli1} and \refeq{eq:pauli2} in  \refeq{eq:pauli3}, that results in
\begin{equation}
  \label{eq:sr7}
  G^p \fan^{-1} G^h =  G^h \fan^{-1} G^p.
\end{equation}
Finally combining \refeq{eq:sr6}, \refeq{eq:sr7} gives an expression for the anomalous part of the self-energy
\begin{align}
  \label{eq:sr8}
  S = \left( G^p \fan^{-1} G^h - \fan \right)^{-1}
\end{align}
We substitute it into the coefficient of the local perturbations $\sim \delta \theta_i^2$ of \refeq{eq:je8} and analyse it in two contributions.
\begin{widetext}
  With \refeq{eq:sr7} the first term immediately reads
  \begin{align}
    \label{eq:sr9}
    G^p S G^h S = \left( G^h \fan^{-1} - F (G^p)^{-1}\right)^{-1} \left( G^p \fan^{-1} - \fan (G^h)^{-1}\right)^{-1}.
  \end{align}
  The second involves a bit more algebra:
  \begin{align}
    \label{eq:sr10}
    \begin{split}
      \fan S \left(1+ \fan S\right) 
      &= \left( G^p \fan^{-1} G^h \fan^{-1} -1\right)^{-1} \left(1+\left( G^p \fan^{-1} G^h \fan^{-1} -1\right)^{-1} \right)\\
      &=\left( G^p \fan^{-1} G^h \fan^{-1} -1\right)^{-1} G^p \fan^{-1} G^h \fan^{-1} \left( G^p \fan^{-1} G^h \fan^{-1} -1\right)^{-1}\\
      &= \left( G^h \fan^{-1} - F (G^p)^{-1}\right)^{-1} \left( G^p \fan^{-1} - \fan (G^h)^{-1}\right)^{-1}.
    \end{split}
  \end{align}
\end{widetext}
It makes the contribution of local phase fluctuations to the variation of the thermodynamic potential vanish (see \refeq{eq:je8}), i.e.
\begin{equation}
  \label{eq:sr8b}
  G^p S G^h S-\fan S- \fan S\fan S = 0
\end{equation}
 and therefore ensures the gauge invariance.

\section{Continuous medium Limit\label{app:tl}}
We take the continuum limit of the Josephson lattice model in order to obtain a relation to the macroscopic observable, the superconducting stiffness $I$. Starting from the long-wavelength approximation
\begin{equation}
  \label{eq:tl1}
  H = \frac{1}{2}\sum_{ij}J_{ij}\theta_{ij}^2
\end{equation}
we assume a rather uniform spatial profile of the low-energy modes. Therefore it is reasonable to interpolate linearly between the plaquettes ($i,j$) as we move them infinitesimally close together and take the continuum-limit
\begin{align}
  \label{eq:tl2}
  \begin{split}
    \theta_{ij} &\to \nabla \theta(r) \left(r-r^\prime\right)\\
    &= \sum_a \frac{\partial \theta}{\partial r_a} \left(r-r^\prime\right)_a.
    \end{split}
\end{align}
In this limit the Hamiltonian reads
\begin{equation}
  \label{eq:tl3}
  H = \frac{1}{2} \sum_{ab} \int d^dr\, \frac{\partial \theta}{\partial r_a} \frac{\partial \theta}{\partial r_b} I_{ab}(r)
\end{equation}
with the $d$-dimensional unit-cell volume $V$ and the superconducting stiffness
\begin{equation}
  \label{eq:tl4}
  I_{ab}(r) = \frac{1}{V^2} \int d^dr^\prime\, J(r - r^\prime) \left(r-r^\prime\right)_a \left(r-r^\prime\right)_b.
\end{equation}
We substitute $R=r-r^\prime$ and insert the Fourier representation of $J$:
\begin{align}
  \label{eq:tl5}
  \begin{split}
    I_{ab} &= \frac{1}{V} \int \frac{d^dq}{\left(2\pi\right)^d} \int d^dR\, e^{iqR} R_a R_b  J(q)\\
    &= -\frac{1}{V} \partial_{q_a} \partial_{q_b} J(q)\Big|_{q=0}
  \end{split}
\end{align}
with
\begin{align}
  \label{eq:tl5b}
  J(q) = \frac{VT}{\left(2\pi\right)^d}\int d^dk\, \Tr_{\omega \alpha} \left(F_k S F_{k-q} S - G^{p\up}_k S G^{h\dn}_{k-q} S\right).
\end{align}
Next we have to evaluate the derivative. After performing the derivative with respect to $q$, we can substitute $k^\prime = k-q$ and perform a partial integration that leads to
\begin{align}
  \label{eq:tl6}
  \begin{split}
    \partial_{q_a} \partial_{q_b} J(q) =& \frac{-VT}{\left(2\pi \right)^d} \int d^dk^\prime \Tr_{\omega \alpha} \Big\{\\
     &\left(\partial_{k^\prime_a} F_{k^\prime -q}\right)S\left(\partial_{k^\prime_b} F_{k^\prime}\right)S\\
    &- \left(\partial_{k^\prime_a} G^{p\up}_{k^\prime -q}\right)S\left(\partial_{k^\prime_b} G^{h\dn}_{k^\prime}\right)S
    \Big\}
  \end{split}
\end{align}
and in \refeq{eq:tl5} finally to
\begin{gather}
  \label{eq:tlstiffaak}
    I_{ab} = \frac{T}{\left(2\pi\right)^d}\int\! d^dk \Tr_{\omega \alpha}\\
    \times \left( \frac{\partial F(k)}{\partial k_a} S \frac{\partial F(k)}{\partial k_b} S
    -\frac{\partial G^{p\up}(k)}{\partial k_a} S \frac{\partial G^{h\dn}(k)}{\partial k_b} S \right)\nonumber
\end{gather}
with the effective Hamiltonian
\begin{equation}
  \label{eq:tlhjosephsoncontinuum}
  H_{\mathrm{eff}}=\frac{1}{2}\sum_{ab} I_{ab}\int d^dr\,\frac{\partial \theta }{\partial r_{a}}\frac{\partial \theta }{\partial r_{b}}.
\end{equation}
Note, that the physical units of the dSC stiffness are restored by
\begin{equation}
  \label{eq:units}
  \begin{split}
    I_\parallel \rightarrow \frac{a_a}{a_b a_c} t\, I_{\parallel},\quad
    I_\perp \rightarrow \frac{a_c}{a_a a_b} t\, I_{\perp}.
  \end{split}
\end{equation}

\begin{figure}
  \centering
  \includegraphics{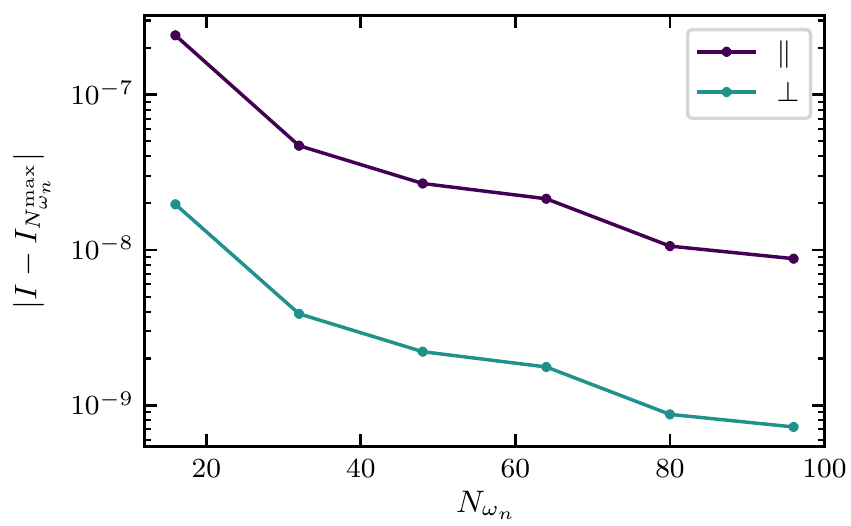}
  \caption{Convergence of the dSC stiffness $I$ with number of Matsubara frequencies $\omega_n$ ($N^\mathrm{max}_{\omega_n} = N_{k} = 128$).}
  \label{fig:stiff_niw}
\end{figure}
\begin{figure}
  \centering
  \includegraphics{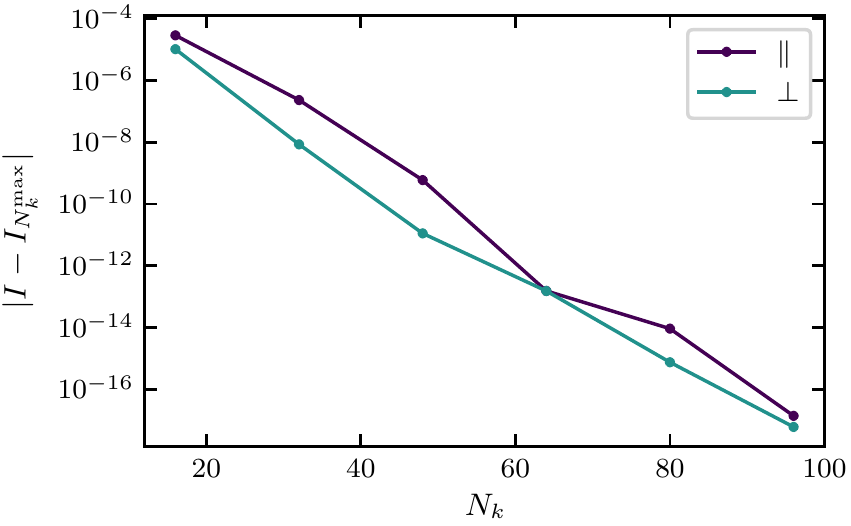}
  \caption{Convergence of the dSC stiffness $I$ with number of $k$-points per dimension ($N^\mathrm{max}_{k} = N_{\omega_n} = 128$).}
  \label{fig:stiff_nk}
\end{figure}
In particular for numerical purposes we express the derivatives in terms of derivatives applied to inverse Green functions
\begin{align}
  \label{eq:tl7}
  \begin{split}
    \partial_{k_a}G &= -G \left(\partial_{k_a} G^{-1} \right) G\\
    %\partial_{k_a} \partial_{k_b} G &= G\left(\partial_{k_a}G^{-1}\right)G\left(\partial_{k_b}G^{-1}\right)G\\
    %&- G\left(\partial_{k_a}\partial_{k_b} G^{-1}\right)G\\
    %&+ G\left(\partial_{k_b}G^{-1}\right)G\left(\partial_{k_a}G^{-1}\right)G,
  \end{split}
\end{align}
since it reduces the differentiation to that of the electron dispersion $G^{-1}(k) \sim t(k)$, that can be performed analytically. Regarding the number of $k$-points per dimension and Matsubara frequencies $\omega_n$ we choose $N_k = N_{i\omega_n} = 64$ which is sufficient for an accuracy of $\sim 10^{-7}$, see \reffig{fig:stiff_niw} and \reffig{fig:stiff_nk}.

\section{London penetration depth\label{app:lpd}}
The London penetration depth describes how far a magnetic field penetrates into the superconductor despite the Meissner effect. The superconductor expells the magnetic field by forming supercurrents. Thereby the magnetic field decays exponentially into the superconductor. In order to describe the Josephson lattice model coupled to an electromagnetic field we start from the gauge-invariant minimal coupling Hamiltonian
\begin{equation}
  \label{eq:lpd1}
  H = \frac{1}{2}\sum_{ab} I_{ab} \int d^dr \left(\frac{\partial \theta}{\partial r} - \frac{e}{\hbar c} 2A\right)_a \left(\frac{\partial \theta}{\partial r} - \frac{e}{\hbar c} 2A\right)_b.
\end{equation}
The factor of ``2'' in front of the gauge field $A$ is essential to ensure gauge invariance. The gauge transformation of the superconducting order parameter $\Phi = \expval{cc}$ is
\begin{equation}
  \label{eq:lpd2}
  c \mapsto c e^{i\frac{e}{\hbar c} \chi},\quad \Phi \mapsto \Phi e^{i\frac{e}{\hbar c} 2\chi},\quad A \mapsto A + \frac{\partial \chi}{\partial r}
\end{equation}
for arbitrary $\chi$. Just as in Landau-Ginzburg theory $\Phi$ can be regarded as the field of the order parameter and its phase we define as $\theta$. According to \refeq{eq:lpd2} $\theta$ transforms under a gauge transformation as $\theta \mapsto 2 e \chi /\hbar c$ and hence \refeq{eq:lpd1} is gauge invariant.

Next we calculate the current given by the derivative of the Hamiltonian with respect to the gauge field
\begin{align}
  \label{eq:lpd3}
  \begin{split}
    j_a &= -c \frac{\partial H}{\partial A_a}\\
    &= \frac{2e}{\hbar}\sum_b I_{ab} \int d^dr \left( \frac{\partial \theta}{\partial r} - \frac{e}{\hbar c} 2A\right)_b,
  \end{split}
\end{align}
absorb $\nabla \theta$ into $A \mapsto A^\prime$ by our choice of gauge
\begin{equation}
  \label{eq:lpd3b}
  j_a = -\frac{2e}{\hbar}\sum_b I_{ab} \int d^dr \frac{e}{\hbar c} 2A^\prime_b,
\end{equation}
and insert it into the Maxwell equation for the current
\begin{equation}
  \label{eq:lpd4}
  \nabla^2 A = - \frac{4\pi}{c} j.
\end{equation}
This gives a differential equation describing the exponential decay of the vector potential into the superconductor
\begin{equation}
  \label{eq:lpd5}
  \nabla^2 A^\prime = \lambda^{-2} A^\prime
\end{equation}
with the penetration depth
\begin{equation}
  \label{eq:lpd6}
  \lambda^{-2} = \frac{16\pi e^2}{\hbar^2 c^2} I.
\end{equation}
Note that both, $I$ and $\lambda$ are matrices in \refeq{eq:lpd6}. Furthermore, \refeq{eq:lpd4} assumes a certain geometry of the setup. The supercurrent $j$ that expells the magnetic field $B = \mathrm{rot} A$ inside the superconductor and $B$ are directed along the main axes of the superconductor. The penetration depth $\lambda$ describes how far the magnetic field or, equivalently, the supercurrent extent into the superconductor. Thus, the direction of the penetration depth is orthogonal to both, that of $j$ and of $B$.

\section{Details of the stiffness dependence on the electronic bandstructure \label{app:ihop}}
\begin{figure}
  \centering
  \includegraphics{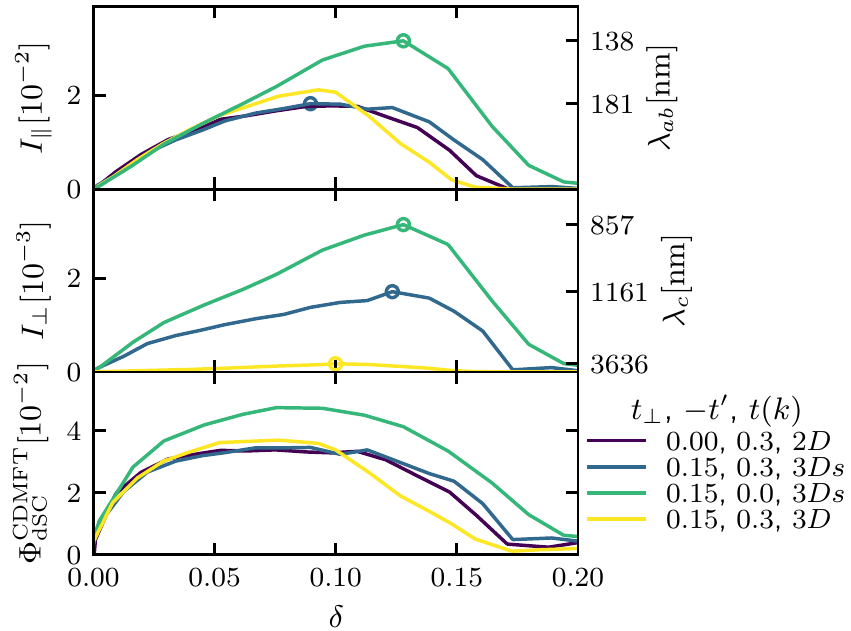}
  \caption{In-plane superconducting stiffness $I_{\parallel}$ (top, left), in-plane penetration depth $\lambda_{ab}$ (top, right), perpendicular superconducting stiffness $I_{\perp}$ (center, left), perpendicular penetration depth (center, right) and CDMFT dSC order parameter $\Phi^{\mathrm{CDMFT}}_{\mathrm{dSC}}$ (bottom) as functions of doping $\delta$ at $T = 1/52 \sim 0.02$. Quantities are shown for different interlayer hoppings $t_\perp$, next-nearest neighbor hoppings $t^\prime$ and also tight-binding lattices $t(k)$.}
  \label{fig:stiff_sco_dop_supp}
\end{figure}
\reffig{fig:stiff_sco_dop_supp} presents the dSC stiffness for all three lattice dispersions. The dSC stiffness of $t^{3D}(k)$ is of similar magnitude as $t^{3Ds}(k)$. In the overdoped regime it is smaller because of the smaller local order parameter $\Phi^{\mathrm{CDMFT}}_{\mathrm{dSC}}$. For the underdoped to optimally doped regimes $t^{3D}(k)$ can be regarded es an effective reduction of $t^\prime$ in terms of the dSC stiffness. In contrast $I_\perp$ is significantly suppressed by the anisotropic interplane model $3D$. Its minimal value of $\lambda_c \sim 3000\nm$ is still in a reasonable range compared to experiments\cite{Homes2004}. Possibly the suppression occurs due to the more pronounced flattness of the $3D$ model's dispersion $t^{3D}(k)$. The derivative of \refeq{eq:stiffaak} is thus much smaller and reduces $I$.

\begin{figure}
  \centering
  \includegraphics{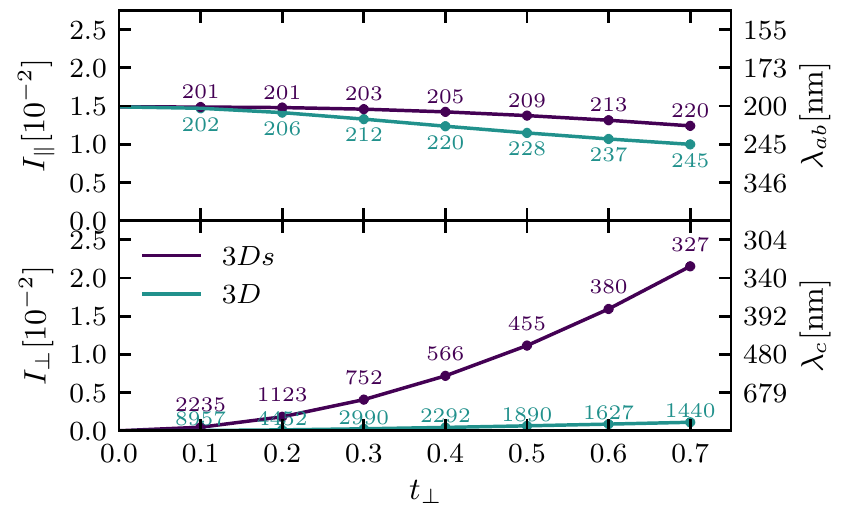}
  \caption{Superconducting stiffness $I_{\parallel / \perp}$ and penetration depth $\lambda_{ab / c}$ as functions of the interlayer hopping $t_\perp$ with in-plane next-nearest neighbor hopping $t^\prime = -0.3$ ($\beta = 52$, $\delta = 0.05$). Results are shown for $3Ds$ and $3D$ lattice dispersions $t(k)$. $t_\perp$ changes only in the Josephson lattice model. The small numbers are values of $\lambda$.}
  \label{fig:stifftperptnnn03}
\end{figure}
Since $I$ can be sensitive to the lattice dispersion it is interesting to examine its dependence on the hopping parameters further. \reffig{fig:stifftperptnnn03} shows $I$ as a function of the interplane hopping $t_\perp$. Both lattice dispersions are considered. It has to be stressed, that for all the data of \reffig{fig:stifftperptnnn03} a single CDMFT calculation is used. The parameters are varied only within the subsequent analysis of the Josephson lattice model. This allows to isolate the effect of the hopping parameters on the phase fluctuations, neglecting the change in the strong-coupling Higgs fluctuations of the plaquette. The CDMFT calculation is performed for the $2D$ lattice and in the underdoped regime ($\delta \sim 0.05$) at cold temperatures ($T\sim 0.02$). This shall reduce a potential bias in the comparison between the $3Ds$ and $3D$ models. For both lattices does $t_\perp$ reduce $I_\parallel$ and increases $I_\perp$. Furthermore the $3D$ model gives smaller $I_{\parallel / \perp}$ for all values of $t_\perp$. In the $3Ds$ lattice $I_\perp$ is more sensitive to $t_\perp$ and in the $3D$ lattice $I_\parallel$ is more sensitive to $t_\perp$.

\begin{figure}
  \centering
  \includegraphics{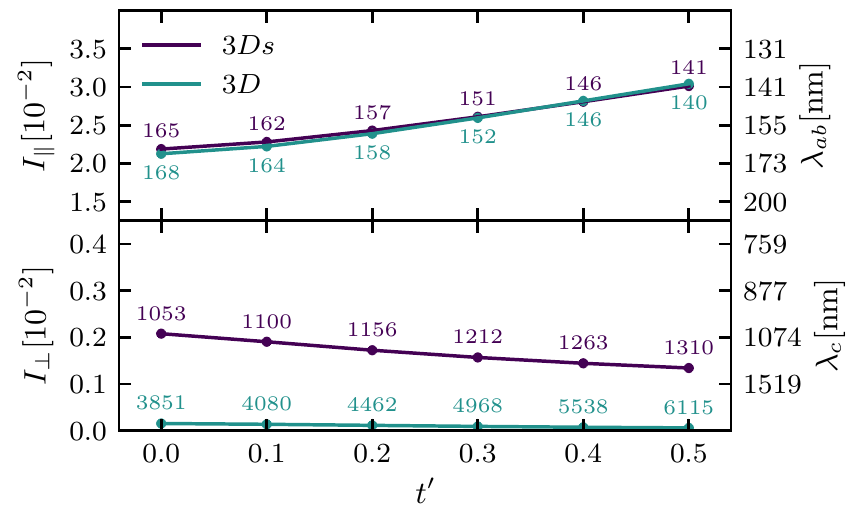}
  \caption{Superconducting stiffness $I_{\parallel / \perp}$ and penetration depth $\lambda_{ab / c}$ as functions of the next-nearest neighbor hopping $t^\prime$ with interlayer hopping $t_\perp = 0.15$ ($\beta = 52$, $\delta = 0.075$). Results are shown for the $3Ds$ lattice dispersion $t(k)$. $t^\prime$ changes only in the Josephson lattice model. The small numbers are values of $\lambda$.}
  \label{fig:stifftnnn}
\end{figure}
A similar analysis is presented in \reffig{fig:stifftnnn}. The single CDMFT calculation is performed at $t^\prime = -0.3$, $\delta\sim 0.075$, $T\sim 0.02$ and $t_\perp = 0.15$ in the $3Ds$ model. Then the subsequent Josephson lattice calculations are done for different in-plane next-nearest neighbor hoppings $t^\prime$. $t^\prime$ has a stronger impact on $I_\parallel$ than on $I_\perp$, which is intuitive as $t^\prime$ and $I_\parallel$ are both in-plane quanitities. Also, in both cases, $3Ds$ and $3D$, $t^\prime$ increases $I_\parallel$ and decreases $I_\perp$. The fact, that it increases $I_\parallel$ is a very interesting trend, because in CDMFT $t^\prime$ diminishes the local order parameter of dSC $\Phi_{\mathrm{dSC}}^{\mathrm{CDMFT}}$. This seems as a contradiction if one interpretes $T^{\mathrm{CDMFT}}_c$ as the $T_c$ of the cuprates\cite{Pavarini2001}, but this is clearly not the case as CDMFT takes into account only spatial correlations within the cluster. It can be speculated based on the $2D$ behavior of $T_{KT}\sim I_\parallel$, that $t^\prime$ has an enhancing effect on the phase fluctuations that are crucial in the underdoped regime and thus increases the critical temperature.

\reffig{fig:stifftperptnnn03} and \reffig{fig:stifftnnn} also allow us to estimate the uncertainty of our predictions on $\lambda$ imposed by the hopping parameters $t_\perp$, $t^\prime$ and to some extent also by the bandstructure. In particular in the case of \ybco it is unclear how well a single band model reflects the bilayer structure. Assuming a one-band model the uncertainty of the correct $t(k)$ and $t_\perp$ translates to an estimated uncertainty of $\Delta\lambda_{ab} \sim 40\nm$ and $\Delta\lambda_c \sim 7500\nm$.

\end{document}